\documentclass[aip,prx,twocolumn,showpacs,superscriptaddress,groupedaddress]{revtex4}  
\usepackage{graphicx}  
\usepackage{dcolumn}   
\usepackage{bm}        
\usepackage{amssymb}   
\usepackage{amsmath}
\usepackage{xfrac}

\usepackage{color}

\preprint{AIP/123-QED}

\begin{document}


\title{Field-Effect Control of Metallic Superconducting Systems}
\author {Federico Paolucci}
\email{federico.paolucci@nano.cnr.it}
\affiliation{INFN Sezione di Pisa, Largo Bruno Pontecorvo, 3, I-56127 Pisa, Italy}
\affiliation{NEST, Instituto Nanoscienze-CNR and Scuola Normale Superiore, I-56127 Pisa, Italy}
\author {Giorgio De Simoni}
\affiliation{NEST, Instituto Nanoscienze-CNR and Scuola Normale Superiore, I-56127 Pisa, Italy}
\author {Paolo Solinas}
\affiliation{CNR SPIN, Via Dodecaneso 33, 16146 Genoa, Italy}
\author {Elia Strambini}
\affiliation{NEST, Instituto Nanoscienze-CNR and Scuola Normale Superiore, I-56127 Pisa, Italy}
\author {Claudio Puglia}
\affiliation{Dipartimento di Fisica dell''Universit\`a di Pisa, Largo Pontecorvo 3, I-56127 Pisa, Italy}
\affiliation{NEST, Instituto Nanoscienze-CNR and Scuola Normale Superiore, I-56127 Pisa, Italy}
\author {Nadia Ligato}
\affiliation{NEST, Instituto Nanoscienze-CNR and Scuola Normale Superiore, I-56127 Pisa, Italy}
\author {Francesco Giazotto}
\affiliation{NEST, Instituto Nanoscienze-CNR and Scuola Normale Superiore, I-56127 Pisa, Italy}

\begin{abstract}
Static electric fields have negligible influence on the electric and transport properties of a metal because of the screening effect.
This belief was extended to conventional metallic superconductors.
However, recent experiments have shown that the superconductor properties can be controlled and manipulated by the application of strong electrostatic fields.
Here, we review the experimental results obtained in the realization of field-effect metallic superconducting devices exploiting this phenomenon. We start by presenting the pioneering results on superconducting Bardeen-Cooper-Schrieffer (BCS) wires and nano-constriction Josephson junctions (Dayem bridges) made of different materials, such as titanium, aluminum and vanadium. 
Then, we show the mastering of the Josephson supercurrent in superconductor-normal metal-superconductor proximity transistors suggesting that the presence of induced superconducting correlations is enough to see this unconventional field-effect. 
Later, we present the control of the interference pattern in a superconducting quantum interference device indicating the \textit{coupling} of the electric field with the superconducting phase. 
We conclude this review discussing some devices that may represent a breakthrough in superconducting quantum and classical computation.
\end{abstract}

\maketitle

\section{Introduction}

The impact of an external static electric field on superconductivity puzzles scientists since about 80 years. 
The first version of the electrodynamic theory of superconductivity of London brothers \cite{London1935} allowed the existence of both static magnetic and electric fields within a superconductor over the London penetration length ($\lambda_L$). 
Experiments by H. London \cite{London1936} and theoretical criticisms by M. von Laue \cite{vonLaue1935} brought slightly after the London brothers to exclude the presence of electrostatic fields into superconductors \cite{London1937}.
Despite a few experiments pointing towards electric field penetration in superconductors \cite{Tao2002,Moro2003}, the surface screening of an external electric field has been always related to the normal metal case \cite{Larkin1963,Lang1970,Ummarino2017,Virtanen2019}, where density functional theory calculations showed maximum penetrations up to $\sim1$ nm in thin polycrystalline films.
Consequently, the effects of an external electrostatic field on superconductivity, \textit{e.g.} the superconducting energy gap ($\Delta$) and critical temperature ($T_C$), has been theoretically discussed for the variation of the surface density of states (DOS), \textit{i.e.} the modulation of the free charge carrier concentration of the superconductor outer layer \cite{Shapiro1984,Burlachkov1993,Lee1996,Lipavsky2006,Morawetz2008}.

In parallel, pioneering gating experiments showed the possibility of tuning the electronic properties of metallic thin films. In particular, electrostatic charge accumulation/depletion have been demonstrated to tune both the conductivity of normal metals \cite{Bonfiglioli1956,Bonfiglioli1959,Berman1975} and the $T_C$ of metallic superconductors \cite{Glover1960,Bonfiglioli1962} of a few percent. 
The breakthrough of tuning the electronic properties of metallic systems has been the implementation of electric double layer transistors (EDLTs) \cite{Locklin2003,Panzer2005}. They rely on polarization of an electrolyte, \textit{i.e.} an ionic conductor and electronic insulator \cite{Kirby2013}, through the application of a voltage between the transistor channel and a counter gate electrode. This technique allows to generate electric fields up to $10^{10}$ V/m and to induce surface charge modulations of $10^{15}$ charges/cm$^2$ \cite{Dhoot2006}. As a consequence, huge modulation of both charge carrier concentration and transport properties have been reported on EDLTs based on low charge concentration 2D material \cite{Misra2007, Shimotani2007,Efetov2010,Gonnelli2015}. In particular, superconductivity has been modulated in layered superconductors \cite{Dhoot2010}, and it has been induced in semiconductors \cite{Ye2009}, topological insulators \cite{Sajadi2018,Fatemi2018} and conventional insulators \cite{Ueno2008}. By contrast, conductivity modulations of about $10\%$ in metal thin films \cite{Daghero2012,Tortello2013} and critical temperature modifications of $\sim1\%$ in metallic superconductors \cite{Choi2014,Piatti2017} have been demonstrated. 

In the last decades no experimental effort has been put in the study of electric field driven modulations of supercurrent in metallic superconductors. The latter have been extensively exploited in low charge concentration superconductors \cite{Fiory1990,Okamoto1992,Mannhart1993,Mannhart1993b} for the realization of superconducting field-effect transistors (SuFETs) \cite{Nishino1989} with interesting applications in superconducting electronics. 

Only in the last two years the impact of electrostatic fields on the critical current ($I_C$) of systems based on conventional metallic superconductors, i.e., described by Bardeen-Cooper-Schrieffer (BCS) theory \cite{Bardeen1957}, has been investigated \cite{DeSimoni2018,Varnava2018,Paolucci2018,Paolucci2019,DeSimoni2019,Paolucci2019_2}. Since these studies take advantage of conventional solid gates, they focus on electric fields reaching the order of $10^8$ V/m and creating negligible variations of surface charge carrier concentration. In particular, ambipolar suppression of $I_C$ has been demonstrated in metallic superconductor films \cite{DeSimoni2018}, nano-constriction Josephson junctions \cite{Paolucci2018,Paolucci2019}, metallic proximized systems \cite{DeSimoni2019}, and fully metallic superconducting quantum interference devices (SQUIDs) \cite{Paolucci2019_2}. 
The unconventional field-effect in metallic superconductor systems could pave the way to the realization of easy fabrication and scalable technologies in the realm of both superconducting electronics and quantum computing.

The purpose of this review is to cover the recent advances on the control of supercurrent in metallic superconductor nano-structures by solid conventional gating. The article is organized as follows: Section \ref{Wires} presents the control of supercurrent in fully metallic field-effect transistors (FETs) made of different Bardeen-Cooper-Schrieffer superconductors (Ti and Al) and fabricated on various substrates (silicon dioxide and sapphire), and the dependence of such effect on the devices lateral dimensions; Section \ref{Josephson} shows the evolution of Josephson effect with the gate bias in monolithic superconducting nano-constrictions (Dayem bridge field-effect transistors, DB-FETs) made of titanium and vanadium; Section \ref{Proximity} describes the field-effect control of Josephson supercurrent in superconductor-normal metal-superconductor transistors (SNS-FETs) made of aluminum and copper; Section \ref{SQUID} presents the control of the interference pattern of a fully metallic SQUID by applying a gate bias on a single Josephson junction (JJ); Section \ref{Applications} depicts some of the possible applications with special attention to classical and quantum computation; and finally Section \ref{Concl} presents a roadmap for future developments both in the direction of intimate understanding of the origin of this unconventional field-effect and for its exploitation for industrial purposes. 

\begin{figure*} [ht!]
	\begin{center}
		\includegraphics [width=1\textwidth]{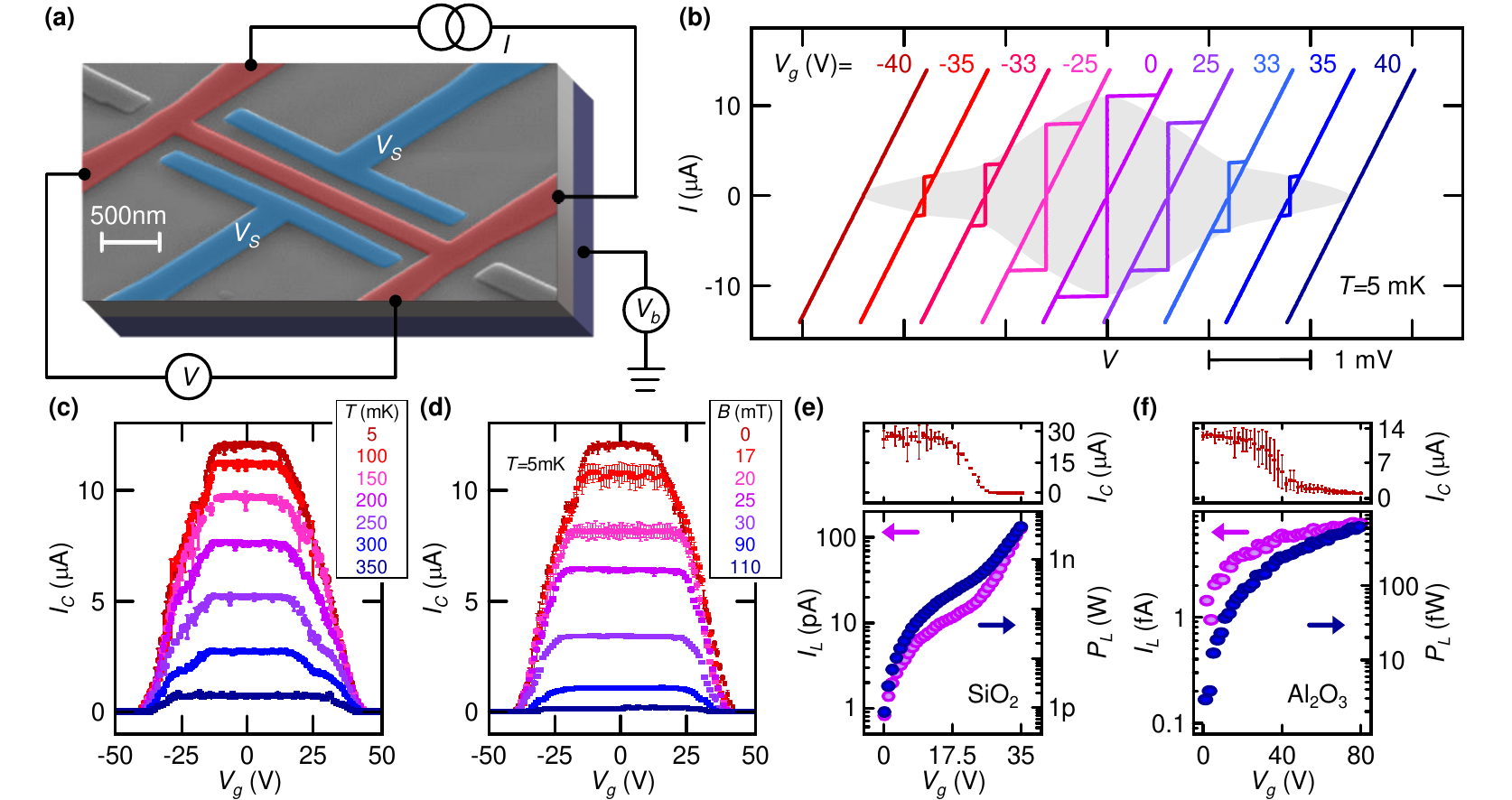}
	\end{center}
	\caption{\label{fig:Fig1}\textbf{Electric field dependence of supercurrent in titanium wires.} (a) A typical Ti-based supercurrent FET. The transistor channel (red) is current biased and the voltage drop is measured in a 4-wire configuration, while the gate voltage can be applied to the back-gate ($V_b$) and the lateral gate ($V_S$) electrodes. (b) Current-voltage characteristics of a titanium supercurrent FET measured at 5 mK (lattice temperature, the electronic temperature is $T_e\simeq 15$ mK) for several values of gate voltage ($V_g=V_b=V_S$). The curves are horizontally offset for clarity. The semitransparent gray area depicts the parameters space where superconductivity persists. (c) Dependence of critical current ($I_C$) on gate voltage ($V_g$) for different values of bath temperature ($T$). (d) Behavior of critical current ($I_C$) versus gate voltage ($V_g$) measured at temperature $T=5$ mK (lattice temperature, the electronic temperature is $T_e\simeq 15$ mK) for different values of perpendicular-to-plane magnetic field ($B$). (e) Top: critical current ($I_C$) versus gate voltage ($V_g$) characteristic measured at $T=32$ mK on a FET fabricated on a silicon dioxide substrate. Bottom: leakage current ($I_L$, left vertical axis, purple dots) and corresponding leakage power ($P_L$, right vertical axis, blue dots) versus gate voltage ($V_g$) for the same device of the top panel. (f) Top: critical current ($I_C$) versus gate voltage ($V_g=V_S$) characteristic measured at $T=32$ mK on a FET fabricated on a sapphire substrate. Bottom: leakage current ($I_L$, left vertical axis, purple dots) and corresponding leakage power ($P_L$, right vertical axis, blue dots) versus gate voltage ($V_g$) for the same device of the top panel.}
\end{figure*}

\section{BCS superconductor thin films and wires}\label{Wires}
This section focuses on the impact of an external electric field on the critical current of conventional BCS metallic superconductors wires and thin films. In particular, this section covers experiments performed on titanium and aluminum based devices fabricated on different substrates (silicon dioxide and sapphire). We show how temperature and magnetic field influence this unconventional gating effect and we discuss how the effect depends on the dimensions of the device.

\subsection{Titanium FETs}\label{Ti}
The titanium supercurrent FETs typically consist of a 200 nm wide and 30 nm thick wire of variable length (ranging from 900 nm to 3 $\mu$m) deposited on a p$^{++}$-doped silicon wafer covered by 300 nm of silicon dioxide \cite{DeSimoni2018}, as displayed in Fig. \ref{fig:Fig1}-a. The device shows a normal state resistance $R_N\simeq45$ $\Omega$ and a critical temperature $T_C \simeq410$ mK. In order to generate the electrostatic fields controlling the supercurrent, two side-gate electrodes (denoted with $V_S$) and a back-gate lead (indicated with $V_b$) have been used. Finally, the critical current is measured by feeding into the wire a current ($I$) and simultaneously recording the voltage drop ($V$).

The dissipationless Cooper pairs transport is highlighted by plotting the current-voltage ($I-V$) characteristics measured at a bath temperature $T=5$ mK (lattice temperature, the electronic temperature is about  $15$ mK), as shown in Fig. \ref{fig:Fig1}-b. The intrinsic wire critical current is $I_C\simeq11$ $\mu$A and shows a hysteretic behavior due to heating induced while switching from the normal to the superconducting state (known as retrapping current) \cite{Courtois2008}. 

The impact of an electrostatic field on $I_C$ is obtained by measuring the $I-V$ characteristic as a function of the gate voltage applied at the same time to all gate electrodes ($V_g=V_S=V_b$) in order to maximize the effect. As depicted in Fig. \ref{fig:Fig1}-b, the critical current lowers by increasing $V_g$ and it is completely suppressed at a critical value $V_g^C\simeq40$ V (resulting in an electric field $E_g^C\sim2\times10^8$ V/m). Differently from SuFETs \cite{Nishino1989}, the electric field does not affect the normal state resistance of the device (the slope of the $I-V$ characteristics is not modified by $V_g$). This behavior can be understood by comparing the intrinsic 2D free charge concentration of a metal ($n_{i,2D}\sim 10^{15}$ cm$^{-2}$ \cite{Ibach}) with its maximum surface modulation due to the applied gate voltage ($\Delta n_{2D}\sim 10^{12}$ cm$^{-2}$). In a depth comparable with the Thomas-Fermi screening length, the maximum relative carrier density modulation is $\Delta n_{2D}/n_{i,2D}\sim 10^{-3}$. As a consequence, the resulting tuning of normal state resistance is expected to be negligibly small (see Fig. \ref{fig:Fig1}-b). Another important difference with respect to SuFETs is that the critical current suppression is symmetric on the polarity of the gate voltage. Since surface charge accumulation/depletion would cause an asymmetric modulation of superconductivity \cite{Fiory1990,Okamoto1992,Mannhart1993,Mannhart1993b}, it cannot account for this unconventional field-effect.

The typical evolution of the gate-induced $I_C$ suppression with temperature is depicted in Fig. \ref{fig:Fig1}-c. At low temperature, the critical current is not affected by low values of $V_g$ and then decreases down to full suppression. By increasing temperature the plateau of $I_C$ with $V_g$ widens, but the critical current full suppression always occurs at the same critical value ($V_g^C(T)\sim$ constant). A microscopic explanation of this effect is still missing, and a phenomenological model \cite{DeSimoni2018} based on the Ginzburg-Landau theory \cite{Ginzburg1950} was shown to grab the dependence of $I_C$ on the applied gate voltage, but the invariance of $V_g^C$ on temperature was an assumption of the model. Finally, Fig. \ref{fig:Fig1}-c shows that the field-effect survives up to $T=350$ mK, that is $\sim85\%$ of the critical temperature of the superconductor. 

\begin{figure} [t!]
\begin{center}
\includegraphics [width=1\columnwidth]{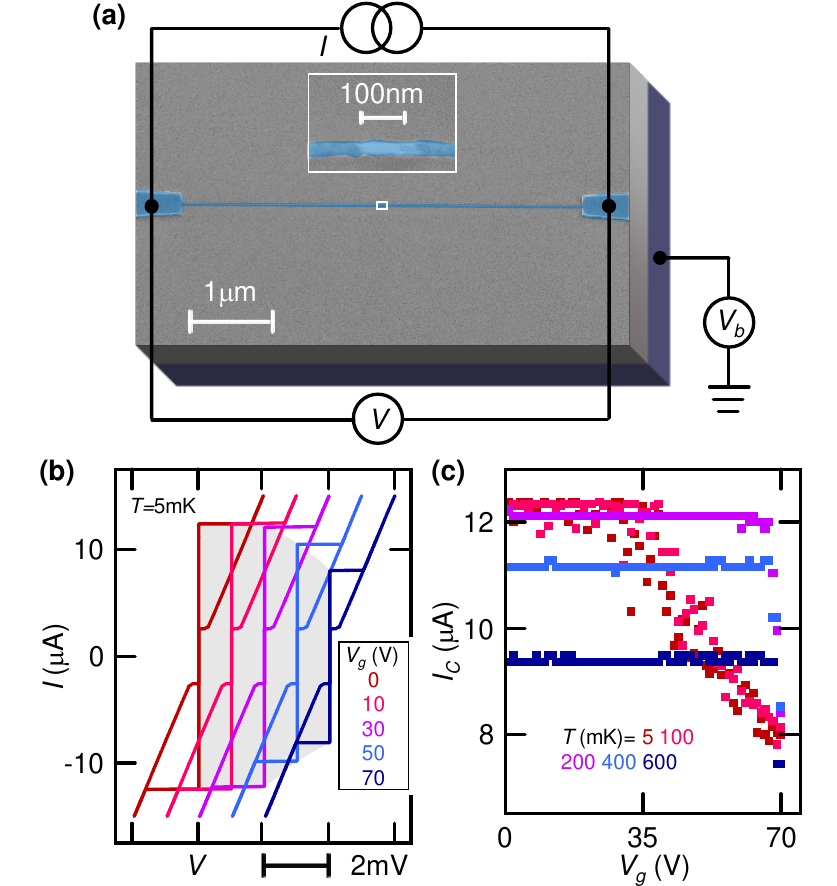}
\end{center}
\caption{\label{fig:Fig2}\textbf{Electric field dependence of supercurrent in aluminum wires.} (a) A typical Al-based supercurrent FET. The transistor channel (blue) is current biased and the voltage drop is measured in a 4-wire configuration, while the gate voltage can be applied to the back-gate ($V_b$). (b) Current-voltage characteristics of an aluminum supercurrent FET measured at 5 mK (lattice temperature, the electronic temperature is $T_e\simeq 15$ mK) for several values of gate voltage ($V_g=V_b$). The curves are horizontally offset for clarity. The semitransparent gray area depicts the parameters space where superconductivity persists. (c) Dependence of the critical current ($I_C$) on gate voltage ($V_g=V_b$) for different values of bath temperature ($T$).}
\end{figure}

Since magnetic field is well known to quench superconductivity \cite{deGennes,tinkham}, the study of the joint impact of electric and magnetic fields on the supercurrent is of extreme interest. Figure \ref{fig:Fig1}-d shows the $I_C$ dependence on $V_g$ for several values of the perpendicular-to-plane magnetic field ($B$) measured at $T=5$ mK (lattice temperature, the electronic temperature is about  $15$ mK). On the one hand, the field-effect remains bipolar in gate voltage and the critical current plateau in $V_g$ widens as in the temperature case. By contrast, the value of the critical voltage weakly decreases by increasing $B$. 
The presented experiments seem to indicate that electric and magnetic fields suppress superconductivity by acting on the same \textit{knob}.

Different experiments have been realized to exclude any possible role of the substrate on the supercurrent suppression. In particular, several devices have been fabricated on undoped silicon covered with 300 nm of thermal silicon dioxide and on sapphire. Both substrates were fully insulating at cryogenic temperatures, therefore exclusively side gate electrodes have been employed. On the one hand, both FETs show a monotonic suppression of the critical current by increasing the gate voltage. The small quantitative variations in the $I_C$ versus $V_G$ characteristics (see Fig. \ref{fig:Fig1}-e and f) can be attributed to differences in the gate/channel distance. On the other hand, the leakage current ($I_L$) and the leakage power ($P_L$) in the two cases differ of several orders of magnitude. Namely, the maximum value of $I_L$ for SiO$_2$ substrate is of $\sim10^{-11}$ A, while for sapphire reaches $\sim10^{-15}$ A. The corresponding leakage power in the entire circuit is $P_L=I_LV_g$ and it is depicted with blue dots in the bottom panels of Fig. \ref{fig:Fig1}-e and f. The typical maximum value of $P_L$ for devices fabricated on silicon dioxide is three orders of magnitude higher than for FETs realized on sapphire ($\sim10^{-10}$ W against $\sim10^{-13}$ W). As a consequence, $I_C$ suppression due to direct injection of hot electrons into the superconducting wire \cite{Morpurgo1998} seems to be excluded.

\subsection{Aluminum FETs}\label{Al}
For both practical and fundamental purposes, it is important to establish if and how the field effect depends on the material used.
To this aim, several experiments on aluminum based devices have been performed \cite{DeSimoni2018}. Figure \ref{fig:Fig2}-a depicts the false color scanning electron micrograph of a typical aluminum FET of length $l=5$ $\mu$m, width $w\simeq30$ nm  and thickness $t=11$ nm. Differently from titanium-based transistors, these devices were fabricated without the possibility to apply a gate bias to side electrodes, thus only a back-gate voltage could control the supercurrent in the wire ($V_g=V_b$). 

The $I-V$ characteristics measured at $T=5$ mK (lattice temperature, the electronic temperature is about  $15$ mK) for several values of $V_g$ are displayed in Figure \ref{fig:Fig2}-b. The traces present the hysteretic behavior typical of diffusive superconducting thin films. In agreement with the results showed for Ti FETs, the critical current is monotonically reduced by increasing the gate voltage  and the normal-state resistance is unaffected by the external electric field. 
However, in this case, the supercurrent is never completely suppressed. The latter seems again to exclude  sizable charge accumulation/depletion at the superconductor surface (such effect would be even more visible on this type of devices since they are thinner than the titanium transistors). The complete dependence of the transistors behavior on temperature is shown in Fig. \ref{fig:Fig2}-c. On the one hand, the critical current suppression persists up to $\sim40\%$ of $T_C\simeq1.2$ K in full agreement with the case of Ti devices. One the other hand, differently from titanium (see Fig. \ref{fig:Fig1}-c), the plateau of $I_C$ dramatically widens for $T\geq200$ mK (the threshold voltage for the beginning of suppression reaches $\sim70$ V). The reason of the difference in the behavior of titanium and aluminum devices is not yet understood.

\subsection{Spatial extension of field-effect}
The study of the spatial extension of field-effect within a conventional metallic superconductor requires special attention for all the different length scales characterizing the physical system: the penetration depth of the electric field at the superconductor surface and the non-local propagation of the condensate perturbation into its bulk. 

Despite the first theory of superconductivity allowed the existence of an electrostatic field inside a superconductor for a depth comparable to the magnetic penetration depth \cite{London1935}, 
more detailed calculations seem to predict an electrostatic screening length ($\lambda_S$) in the superconductor similar to the one of a normal metal \cite{Larkin1963,Lang1970,Ummarino2017, Virtanen2019}. 
The experimental study of field-effect on NbN \cite{Piatti2017} and theoretical calculations on Pb \cite{Ummarino2017} showed a screening length dependent on the strength of the applied electric field. In particular, the obtained penetration depth ranges from $\lambda_S\simeq\lambda_{Th}$ for low values of external electric field ($10^{8}$ V/m) to $\lambda_S\simeq5\lambda_{Th}$ for $E\sim10^{10}$ V/m \cite{Ummarino2017,Piatti2017}, where $\lambda_{Th}\sim0.1$ nm is the Thomas-Fermi length in a normal metal \cite{Ibach}. 

At the same time, the Cooper pairs of the condensate are separated by one coherence length ($\xi_0$) and we would expect that the perturbation of the correlated electrons on the surface is felt over such length-scales.
This seems to be the case for the charge accumulation/depletion effect \cite{Ummarino2017} for which it has been theoretically shown that, at $T\sim T_C$, the field-effect propagates into the superconductor for at least $\xi_0$.
In fact, the electric field perturbed surface couples to the superconductor bulk via the superconducting proximity effect \cite{Holm1932}. These calculations exploited the one band s-wave Eliashberg equations \cite{Carbotte1990} and result in a complex response of the system to the external static field that takes into account both $\lambda_S$ and the spatial dimensions of the device. 

\begin{figure} [t!]
	\begin{center}
		\includegraphics [width=1\columnwidth]{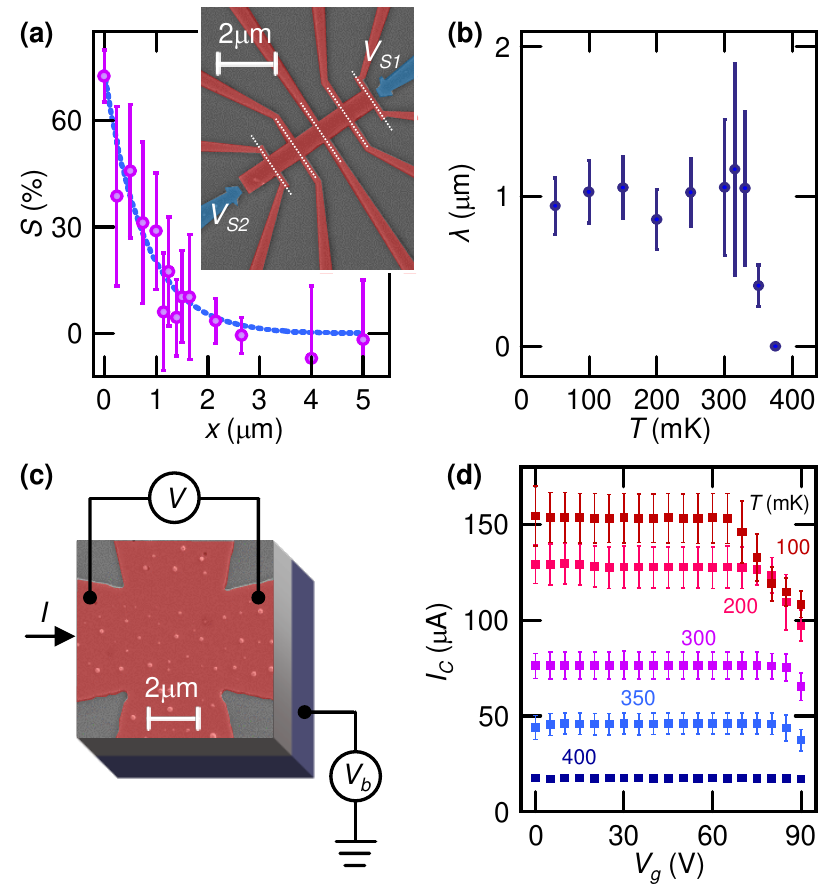}
	\end{center}
	\caption{\label{fig:Fig3}\textbf{Dimension-dependence of the critical current suppression.} (a) Critical current suppression parameter ($S$) versus distance $x$ measured at 5 mK (lattice temperature, the electronic temperature is $T_e\simeq 15$ mK). The dashed blue line is an exponential decay fit. Inset: pseudo-color scanning electron micrograph of a typical Ti comb-like FET used to investigate the spatial extension of the electric field-induced critical current suppression. The core of the device is represented in red, while the two side gate electrodes ($V_{S1}$ and $V_{S2}$) in blue. The white dashed lines indicate the direction of the measured critical current. (b) Behavior of the electrostatic attenuation length ($\lambda$) versus bath temperature ($T$). (c) A typical Ti-based 2D FET. The transistor channel (red) is current biased and the voltage drop is measured in 4-wire configuration, while the gate voltage can be applied to the back-gate ($V_b$). (d) Dependence of critical current ($I_C$) on gate voltage ($V_g=V_b$) for different values of bath temperature ($T$).}
\end{figure}

The typical titanium coherence length extracted from the devices shown in Section \ref{Ti} is $\xi_{0,\rm{Ti}}\simeq100$ nm. In the discussed experiments, the dimensions of the transistors perpendicular to the applied electric field were comparable with the coherence length, namely the thickness was $\sim0.3\xi_{0,\rm{Ti}}$ while the width was $\sim2\xi_{0,\rm{Ti}}$. 
Under these conditions, the perturbation induced by the electric field on the surface affects the whole device.
However, for a more quantitative analysis of the spatial extension of the electric field-induced non-local effect in the superconductor, a different \textit{ad hoc} geometry has been fabricated and used.

The inset of Fig. \ref{fig:Fig3}-a shows a typical comb-like device which allowed to extract the spatial extension of the non-local electric field-effect in titanium. In particular, the $I_C$ versus $V_g$ characteristics have been measured along the wires indicated by the white lines. At each position it was possible to determine the critical current suppression parameter $S=100\times[I_C(V_{Si}=0)-I_C(V_{Si}=90\text{V})]/I_C(V_{Si}=0)$, where $i=1,2$ represents the used gate electrode ($V_{S1}$ or $V_{S2}$). Figure \ref{fig:Fig3}-a shows the parameter $S$ as a function of the distance between each wire and the lateral edge of the comb structure ($x$). The effect of the electric field exponentially vanishes by moving towards the superconductor bulk with an attenuation length $\lambda\simeq770\pm150$ nm (the fit is represented by the dashed blue line in Fig. \ref{fig:Fig3}-a). Interestingly, this value is several times the coherence length ($\sim7$) and in reasonable agreement with the London penetration depth extracted from the devices shown in Section \ref{Ti} ($\lambda_{L,Ti}\simeq900$ nm).

\begin{figure*} [ht!]
	\begin{center}
		\includegraphics [width=1\textwidth]{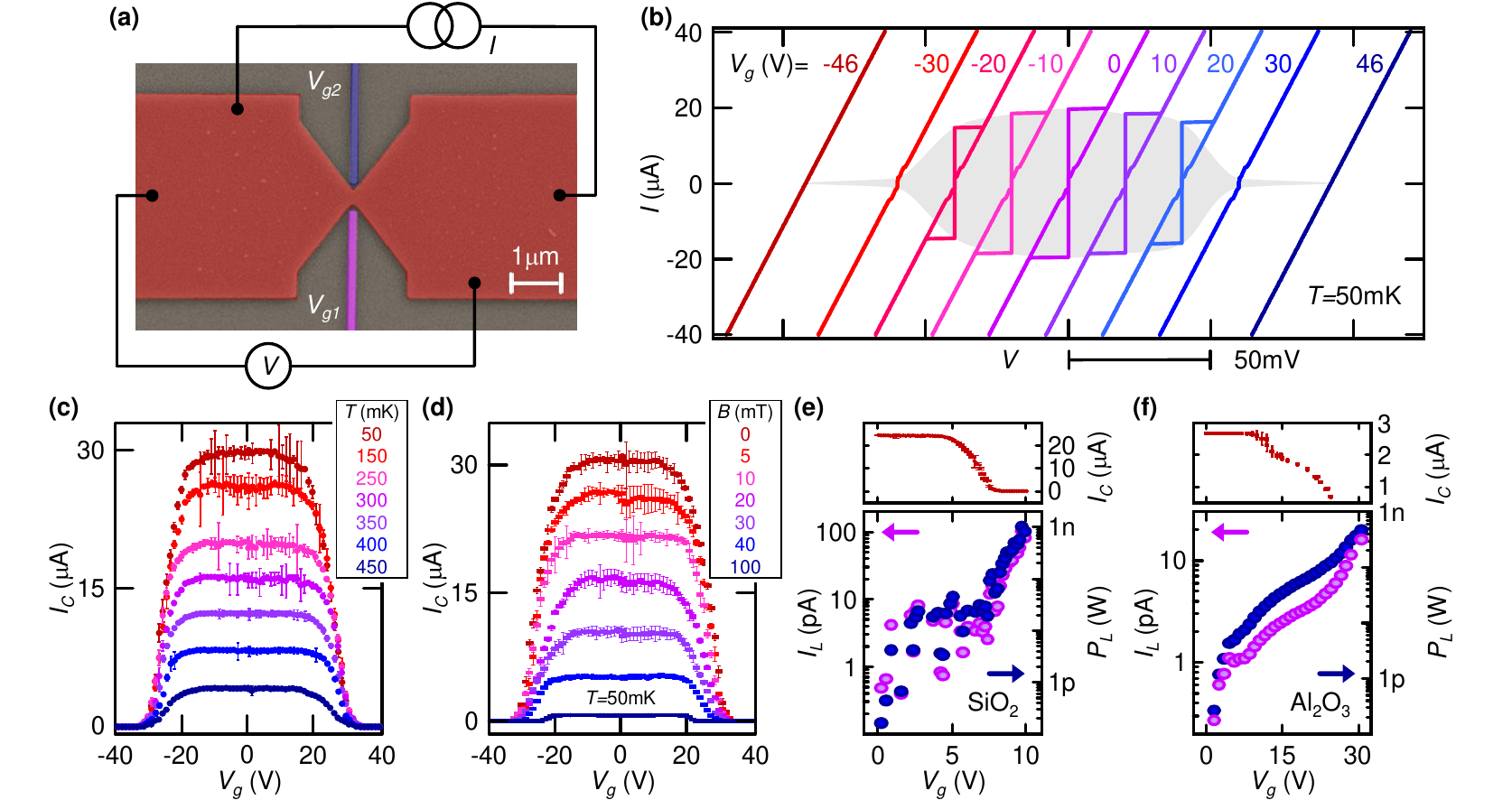}
	\end{center}
	\caption{\label{fig:Fig4}\textbf{Electric field dependence of supercurrent in titanium DB-FETs.} (a) False color scanning electron micrograph of a typical Ti-based DB-FET. The transistor channel (red) is current biased and the voltage drop is measured in 4-wire configuration, while the gate voltage can be independently applied to the lateral gate electrodes ($V_{g1}$ and $V_{g2}$). (b) Current-voltage characteristics of a titanium DB-FET measured at 50 mK for several values of gate voltage ($V_g=V_{g1}=V_{g2}$). The curves are horizontally offset proportionally to $V_g$ for clarity. The semitransparent gray area depicts the parameters space where superconductivity persists. (c) Dependence of critical current ($I_C$) on gate voltage ($V_g$) for different values of bath temperature ($T$). (d) Behavior of critical current ($I_C$) versus gate voltage ($V_g$) measured at temperature $T=50$ mK for different values of perpendicular-to-plane magnetic field ($B$). (e) Top: critical current ($I_C$) versus gate voltage ($V_g=V_{g1}=V_{g2}$) characteristic measured at $T=50$ mK on a different DB-FET fabricated on a silicon dioxide substrate. Bottom: leakage current ($I_L$, left vertical axis, purple dots) and corresponding leakage power ($P_L$, right vertical axis, blue dots) versus gate voltage ($V_g$) for the same device of the top panel. (f) Top: critical current ($I_C$) versus gate voltage ($V_g=V_{g1}=V_{g2}$) characteristic measured at $T=50$ mK on a FET fabricated on a sapphire substrate. Bottom: leakage current ($I_L$, left vertical axis, purple dots) and corresponding leakage power ($P_L$, right vertical axis, blue dots) versus gate voltage ($V_g$) for the same device of the top panel.}
\end{figure*}

The same experiments have been repeated for different bath temperatures. The resulting temperature dependence of the spatial attenuation length is shown in Fig. \ref{fig:Fig3}-b. The parameter $S$ is almost constant for $T\leq330$ mK (about $80\%$ of $T_C$) and it rapidly decays to full quench around 375 mK. This behavior seems to indicate that the number of thermally activated quasiparticles (increasing dramatically near $T_C$ \cite{tinkham}) strongly influences the spatial extension of field-effect. Furthermore, the attenuation length obtained in these experiments is much shorter than the typical thermal relaxation length in superconductors \cite{Fornieri2017} (several tens of micrometers). So, once more, direct hot electron injection can be excluded.

In order to study the field-effect on a superconducting region of lateral dimension larger than both the coherence length and the London penetration depth, a square-shaped titanium transistor shown in Fig. \ref{fig:Fig3}-c has been realized. The FET consists of a 30-nm-thick, 4-$\mu$m-long and 4-$\mu$m-wide channel with the possibility of exclusively apply a back-gate bias. The $I_C$ versus $V_g$ characteristics measured at different values of bath temperature (see Fig. \ref{fig:Fig3}-d) show a behavior similar to Ti wires. Therefore, the direction parallel to the applied electrostatic field appears to be the only relevant spatial scale for this peculiar field-effect. 

\section{Nano-constriction Josephson junctions}\label{Josephson}
This section reports gating experiments on nano-constriction Josephson junctions (Dayem bridges) made of conventional BCS metallic superconductors. In particular, we cover studies on Dayem-Bridge Field-Effect Transistors (DB-FETs) based on titanium and vanadium thin films and fabricated on different substrates (silicon dioxide and sapphire). First, we show how temperature and magnetic field influence the electric field control of supercurrent in the DB-FETs. Then, we show the independence of the critical temperature of the device on gate voltage and how Josephson effect depends on gating. Finally, we discuss the combined effect of two independently controlled gate electrodes on the DB-FET critical current. 

\begin{figure} [t!]
\begin{center}
\includegraphics [width=1\columnwidth]{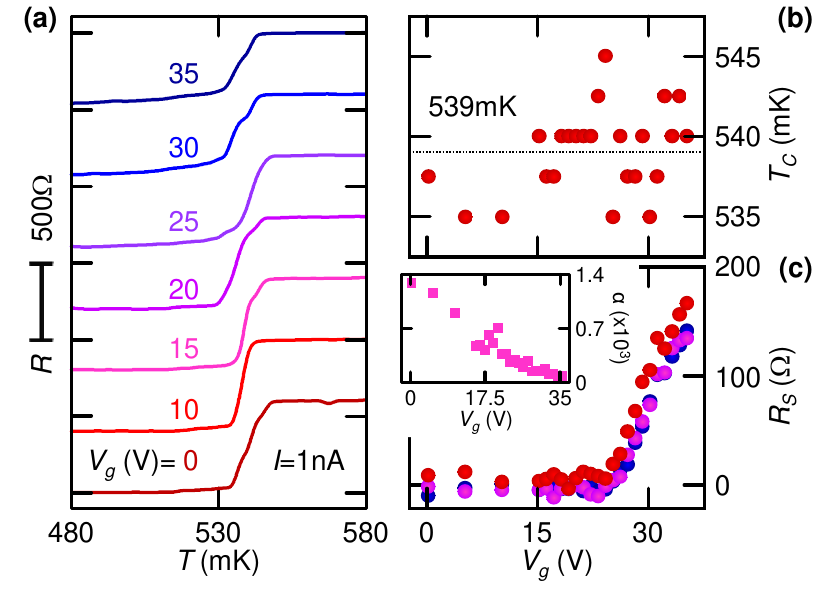}
\end{center}
\caption{\label{fig:Fig5}\textbf{Electric field dependence of critical temperature in titanium DB-FETs.} (a) Resistance ($R$) versus temperature ($T$) characteristics measured at $I=1$ nA for different values of gate voltage ($V_g=V_{g1}=V_{g2}$). (b) Critical temperature ($T_C$) as a function of gate voltage ($V_g$) extracted from the data in (a). The dotted black line represents the average critical temperature $T_C=539$ mK. (c) Resistance in the superconducting state ($R_S$) as a function of gate voltage ($V_g$) at three different temperature: 480 mK (red), 500 mK (pink) and 520 mK (blue). Inset: electrothermal parameter ($\alpha$) as a function of gate voltage ($V_g$) extracted from the data in (a).}
\end{figure}

\subsection{Titanium DB-FETs}\label{DBTi}

\begin{figure*} [ht!]
\begin{center}
\includegraphics [width=1\textwidth]{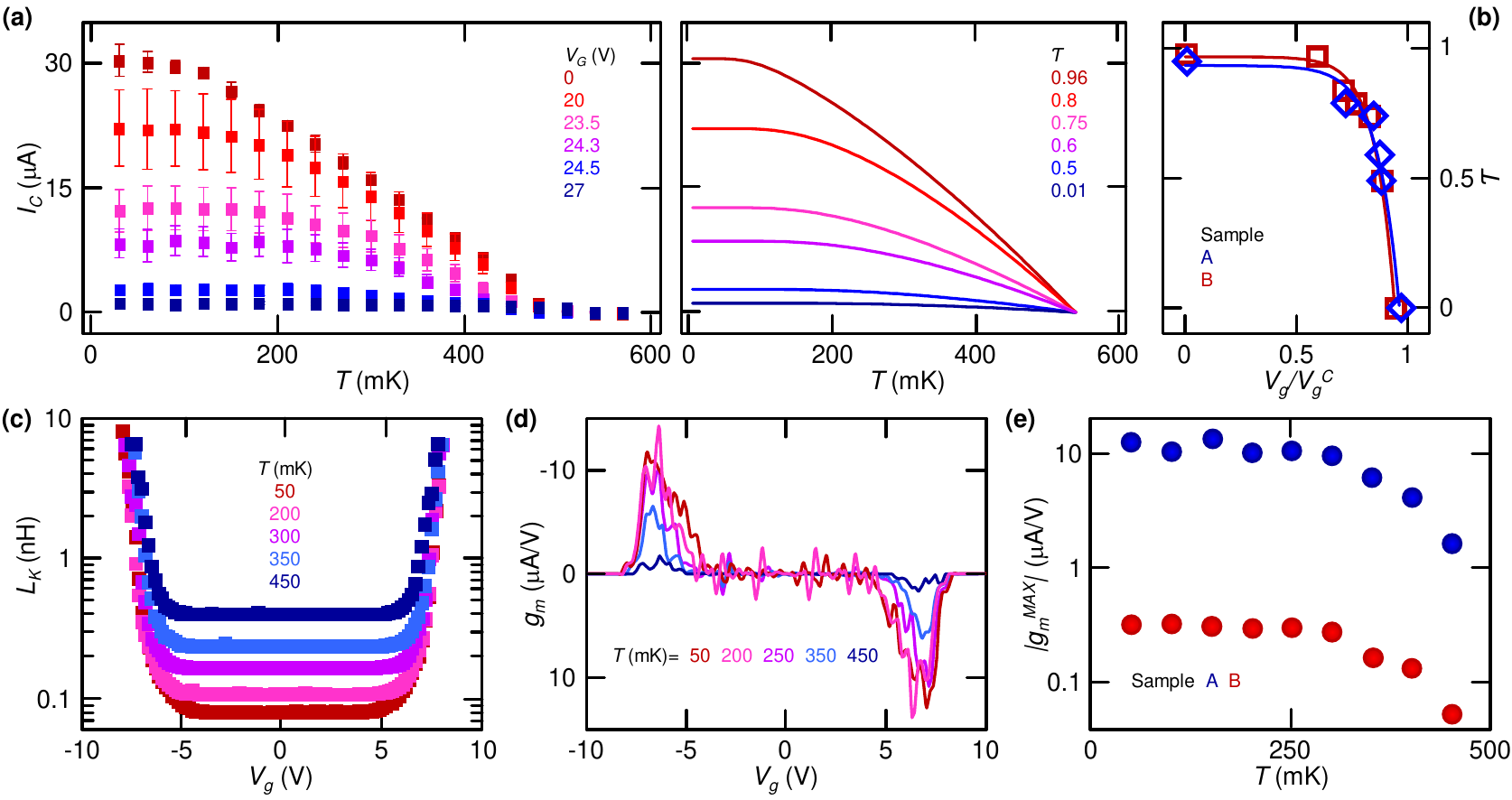}
\end{center}
\caption{\label{fig:Fig6}\textbf{Electric field dependence of Josephson effect in DB-FETs.} (a) Left panel: dependence of critical current ($I_C$) on temperature ($T$) for different values of gate voltage ($V_g=V_{g1}=V_{g2}$). Right panel: fit of $I_C$ versus $T$ calculated with the Kulik-Omelyanchuk theory generalized for an arbitrary effective transmissivity $\mathcal{T}$. The resulting values of $\mathcal{T}$ correspond to the gate voltages represented with the same color in left panel. (b) Effective transmissivity of the Josephson junction $\mathcal{T}$ versus normalized gate voltage ($V_g/V_{g}^C$) extracted from two different devices. (c) Josephson kinetic inductance ($L_K$) versus gate voltage ($V_g=V_{g1}=V_{g2}$) of a typical DB-FET for different values of temperature ($T$). (d) Transconductance ($G_m$) as a function of gate voltage for the same device in (c) for different values of temperature ($T$). (e) Absolute value of the maximum transconductance ($g_m^{MAX}$) as a function of temperature ($T$) for two different DB-FETs.}
\end{figure*}

A typical titanium DB-FET consists of a nano-constriction (about $125$ nm long and $300$ nm wide) interrupting an approximately $4$-$\mu$m-wide and $30$-nm-thick Ti strip, as shown by the false color scanning electron micrograph in Fig. \ref{fig:Fig4}-a. 
The resulting normal state resistance is $R_N\simeq600$ $\Omega$ with an unprecedented critical temperature of about $540~$mK \cite{Paolucci2018,Paolucci2019}. The latter has been obtained by optimal growth conditions of the Ti thin film \cite{Paolucci2018}. Two independent gate voltages ($V_{g1}$ and $V_{g2}$) can be applied on the JJ region through two side gate electrodes (the violet and blue leads, respectively) placed at a distance ranging between $80~$nm and $120~$nm from the constriction. 

The $I-V$ characteristics for different values of gate voltage ($V_g=V_{g1}=V_{g2}$) at a bath temperature $T=50$ mK allows us to determine the performances of the DB-FETs (see Fig. \ref{fig:Fig4}-b). At $V_g=0$ the current versus voltage characteristic is hysteretic owing to quasiparticles overheating while switching from the normal to the superconducting state \cite{Courtois2008}. In particular, we measured a critical current $I_C\simeq18$ $\mu$A and a retrapping current $I_C\simeq1.6$ $\mu$A. By increasing $V_g$ the critical current is monotonically suppressed until its full quenching at $V_g=46$ V. On the contrary, the retrapping current remains almost constant until $I_C(V_g)>I_R(0)$,  and it acquires the same value of the critical current [$I_R(V_g)=I_C(V_g)$] by further increasing the gate voltage. The field-effect driven suppression of the critical current is symmetric in the polarity of the gate voltage. Furthermore, the normal state resistance of the DB-FETs is completely unaffected by the applied gate bias. These results are in full agreement with the experiments presented in Section \ref{Wires} for superconducting wires \cite{DeSimoni2018} and in contrast with the typical phenomenology of Josephson field-effect transistors (JoFETs) \cite{Takayanagi1985,Akazaki1996,Doh2005,Xiang2006}. As a consequence, charge accumulation/depletion has been again excluded as possible source of the presented $I_C$ suppression \cite{Paolucci2018}.

The complete behavior of a DB-FET is resumed in Fig. \ref{fig:Fig4}-c, where the critical current ($I_C$) is plotted versus gate voltage ($V_g=V_{g1}=V_{g2}$) for different values of temperature in the range $50-450~$mK. At $T=50~$ mK, the critical current is almost unaffected for values of gate voltage lower than about 15 V and, then, it starts to drop by further increasing $V_g$ down to full suppression at the critical value $V_g^{C}\simeq32$ V. For higher values of temperature, the $I_C$ plateau widens, but the value of $V_g^{C}$ stays almost constant in the complete temperature range (up to about $0.85T_C$). 

The combined impact of electric and magnetic fields on the critical current in the DB-FETs is summarized in Fig. \ref{fig:Fig4}-d, where the dependence of $I_C$ on $V_g$ is shown for different values of the perpendicular-to-plane magnetic field. Interestingly, the critical current plateau widens by increasing the magnetic field, but the value of $V_g^C$ decreases by rising $B$. 
The widening of the plateau of constant $I_C$ for increasing magnetic field seems to further exclude mere heating as the origin of the effect, because the critical current drops faster with $B$ while approaching the critical temperature \cite{Mydosh1965}.

Analogously to supercurrent FETs \cite{DeSimoni2018}, titanium based DB-FETs have been realized on different insulating substrates to exclude any role of the substrate (such as the leakage current). In particular, several devices have been fabricated on 300 nm intrinsic thermal silicon dioxide (see Fig. \ref{fig:Fig4}-e) and sapphire (see Fig. \ref{fig:Fig4}-f). Typically, the DB-FETs fabricated on sapphire show a lower leakage current ($I_L$) and a lower injected power ($P_L$) than in the case of silicon dioxide substrate. Different devices made on silicon dioxide substrate show complete suppression of $I_C$ at completely different values of leakage current and power \cite{Paolucci2018}. As a consequence, the $I_C$ suppression is unlikely to be ascribed to direct current injection driven quasiparticle overheating.

\subsection{Critical temperature studies in DB-FETs}\label{DBTc}
The study of the resistance ($R$) versus temperature ($T$) characteristics of the DB-FET for different value of gate voltage ($V_g=V_{g1}=V_{g2}$) gives interesting information about the mechanism generating the supercurrent suppression (see Fig. \ref{fig:Fig5}-a for the raw data). First of all, Fig. \ref{fig:Fig5}-b shows that the critical temperature stays constant over the complete range of applied values of gate voltage \cite{Paolucci2018,Paolucci2019}. For instance, a reduction of $\sim80\%$ of $I_C$ has been measured for $V_g=25$ V (see Fig. \ref{fig:Fig4}-c), while the critical temperature was completely unaffected. Since any heat injection reducing the critical current would lower the critical temperature \cite{Morpurgo1998}, a role of hot carrier injection can be excluded. 

The normal-state resistance is unaffected by the gate bias. Its value $R_N\simeq 600$ $\Omega$ is comparable with the values extracted from the current versus voltage characteristics shown in Fig. \ref{fig:Fig4}-b. These results are in full agreement with the data reported for superconducting wires \cite{DeSimoni2018}. Therefore, once more, charge accumulation/depletion has been considered negligible \cite{Paolucci2018,Paolucci2019}.

At the same time, the resistance in the superconducting state ($R_S$) strongly depends on $V_g$ near full supercurrent suppression (see Fig. \ref{fig:Fig5}-c). This can be attributed to the creation of a subgap dissipative component \cite{Paolucci2019} due to the action of the external electrostatic field. In particular, this component could arise from the creation of an inhomogeneous mixed state composed of superconducting and normal metal puddles \cite{Paolucci2018}. In this view, the critical current of the DB-FET would be strongly suppressed by $V_g$, but the critical temperature remains constant.

A good figure of merit for the sharpness of the superconducting transition is the \textit{electrothermal} parameter $\alpha=[T/T(T,I)]/[dR(T,I)/dt]$, where $R$ is the resistance, $I$ is the current and $T$ is the temperature \cite{Harwin2017}. The dependence of $\alpha$ calculated at half of the superconductor/normal metal transition is shown in the inset of Fig. \ref{fig:Fig5}-c. The electrothermal parameter decreases by increasing $V_g$, because the superconducting transition decreases in height and  becomes wider in temperature. At the inset of the $I_C$ suppression, the electric field seems to create instabilities in the Cooper condensate (see the noise in Fig. \ref{fig:Fig4}-c). As a consequence, the superconducting transition is randomized and the parameter $\alpha$ shows strong fluctuations with the gate voltage. 

\subsection{Josephson effect studies in DB-FETs}\label{DBTc}
The dependence of the Josephson coupling on the gate voltage in the DB-FET provided other interesting insights about this unconventional field-effect \cite{Paolucci2019}. 
The behavior of $I_C$ with temperature ($T$) recorded for different values of gate voltage is shown in the left panel of Fig. \ref{fig:Fig6}-a. 
The critical current is monotonically damped by increasing $V_g=V_{g1}=V_{g2}$ in all the studied temperature range, but the temperature of full suppression of the Josephson supercurrent is always the intrinsic value of the Ti thin film, that is $T_C\simeq540$ mK.
Since quasiparticles overheating would cause an increase of the effective electronic temperature ($T_{eff}$) \cite{Morpurgo1998}, the Josephson supercurrent would vanish when $T_{eff}=T_C$, but $T<T_{eff}$. Therefore, these measurements seem to exclude again a hot electron injection mechanism.

The experimental data can be fit with a theoretical equation which describes the critical current of a Josephson constriction in the short-junction limit \cite{Golubov2004}

\begin{equation}
\begin{split}
I_C(\varphi)=\frac{\pi\Delta}{2eR_N}\frac{\sin\varphi}{\sqrt{1-\mathcal{T}\sin^2\frac{\varphi}{2}}}\\
\times\tanh\left[ \frac{\Delta}{2T}\sqrt{1-\mathcal{T}\sin^2\frac{\varphi}{2}}\right].
\end{split}
\label{eq:CritCurr}
\end{equation}
Here, $\varphi$ is the phase difference across the constriction, $\Delta$ is the pairing potential, $e$ is the electron charge and $\mathcal{T}$ is the transmission probability through the junction. The normal state resistance can be written as $R_N=\sfrac{R_{Sh}}{\mathcal{T}}=\sfrac{(4\pi^2\hbar)}{(e^2k_F^2S\mathcal{T}})$, with $R_{Sh}$ the Sharvin resistance, $k_F$ the Fermi wave vector and $S$ the constriction cross-sectional area. If $\mathcal{T}=1$, Eq. \ref{eq:CritCurr} reduces to the Kulik-Omelyanchuck advanced theory for a ballistic clean constriction \cite{Kulik1977}. On the other hand, for $\mathcal{T}\ll1$ it reduces to the Ambegaokar-Baratoff model \cite{Ambegaokar1963} for a tunnel Josephson junction. 

The fits of the experimental data are shown in the right panel of Fig. \ref{fig:Fig6}-a, where the values of $\mathcal{T}$ correspond to the values of gate voltage represented with the same color in the left panel.
In the experiment $R_N$ is constant with respect to $V_g$, therefore in this model we assume that only some channels do not contribute to the supercurrent flow \cite{Paolucci2019}. 
Within this hypothesis, the model produced curves in good agreement with the experimental data. In particular, the effective transmission probability seems to decrease by increasing gate voltage. This means that the behavior of the DB-FET goes from the typical Kulik-Omelyanchuk characteristic of a clean ballistic constriction to the Ambegaokar-Baratoff temperature dependence of a tunnel $JJ$.

Figure \ref{fig:Fig6}-b shows the comparison of the behavior of $\mathcal{T}$ with the gate voltage for two devices (labeled as $A$ and $B$). Since the gate electrodes were placed at different distances from the Josephson junction, the gate voltage has been normalized with respect to the value for complete $I_C$ suppression ($V_g/V_{g}^C$). Although the two devices showed completely different values of $V_{g}^C$, the dependence of $\mathcal{T}$ on the gate voltage seems to be universal. In particular, the effective transmission probability drops dramatically when the gate voltage approaches its critical value. In agreement with the dissipative states growing in the $I-V$ characteristics \cite{Paolucci2019}, this result suggests the formation of superconducting/normal metal disordered regions with a non-continuous dissipationless path.

The kinetic inductance is the first physical quantity typically analyzed in Josephson junctions and plays a fundamental role in gate-tunable transmons qubits \cite{Larsen2015,deLange2015}. It is defined as $L_K=\hbar / (2eI_C)$, where $\hbar$ is the reduced Planck constant. Figure \ref{fig:Fig6}-c shows the $L_K$ vs $V_g$ characteristics extracted from the measurements carried out on sample $A$ at several values of bath temperature. 
The maximum value of kinetic inductance is $L_K\simeq8$ nH, while gate-dependent modulations from hundreds of fH to tens of nH have been obtained.

The transconductance is one of the most used figures of merit for field-effect transistors. In superconducting FETs, it is defined as the variation of critical current with the applied gate bias ($g_m=dI_C/dV_g$). The dependence of $g_m$ with $V_g$ extracted from the measurements carried out on sample $A$ at several values of bath temperature is shown in Fig. \ref{fig:Fig6}-d. To highlight the stark temperature dependence of the transconductance, the absolute value of its maximum ($|g_m^{MAX}|$) is plotted as a function of temperature for two different devices (see Fig. \ref{fig:Fig6}-e). In particular, $|g_m^{MAX}|$ stays almost constant up to $\sim0.5T_C$ and then decreases of almost one order of magnitude while approaching the critical temperature. The maximum values of transconductance measured in the DB-FETs ($\sim 15$ $\mu$A in sample $A$) is comparable with that obtained in thin-film JoFETs \cite{Akazaki1996} and several orders of magnitude higher than that of nanowire Josephson transistors \cite{Doh2005}.

\begin{figure} [t!]
	\begin{center}
		\includegraphics [width=1\columnwidth]{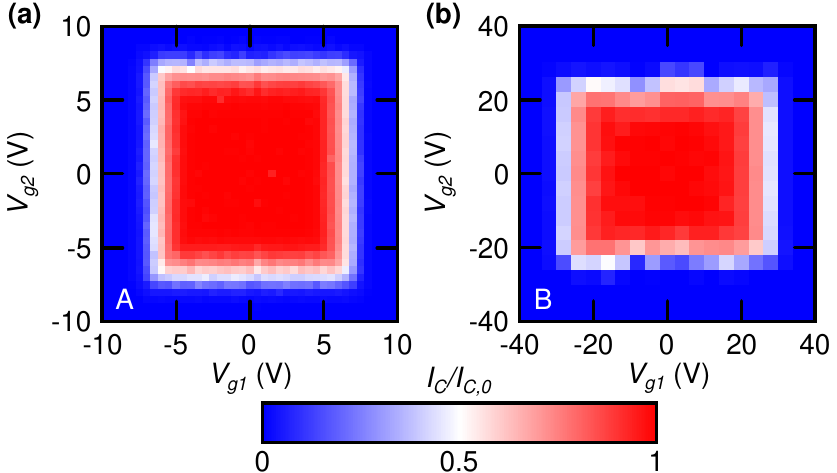}
	\end{center}
	\caption{\label{fig:Fig7}\textbf{Combined effect of two electric fields on titanium Dayem bridges.} (a), (b) Color plot of the normalized critical current as a function of $V_{g1}$ (x-axis) and $V_{g2}$ (y-axis) for two different devices ($A$ and $B$).}
\end{figure}

\subsection{Impact of two independent electric fields}\label{DB2Gate}
The independent use of the two side gate electrodes placed at the opposite flanks of the JJ (see Fig. \ref{fig:Fig4}-a) can give a physical insight about the spatial extension of field-effect. The normalized critical current ($I_C/I_{C,0}$, where $I_{C,0}$ is the intrinsic critical current) as a function of both $V_{g1}$ and $V_{g2}$ is shown in Fig. \ref{fig:Fig7}-a,b for samples $A$ and $B$, respectively. 
The square-like shape indicates the independence of the effect of the two electric field on the critical current. 
This seems to suggest that the gate-induced suppression of critical current is related to a surface effect, which affects non-locally superconductivity. 
In agreement with previous experiments \cite{DeSimoni2018} and calculations \cite{Ummarino2017}, this surface perturbation may affect the superconductor for a depth of a few times the superconducting coherence length ($\xi$), that is the characteristic length scale for superconductivity.

The quantitative difference in the values of $V_{g}^C$ between the two devices has been attributed to the variance in the distance between the gate electrodes and the JJs ($\sim80$ nm in sample $A$ and $\sim120$ nm in sample $B$). 
Furthermore, the rectangular shape in Fig. \ref{fig:Fig7}-b has been attributed to the different distance of $V_{g1}$ and $V_{g2}$ from the active region in DB-FET $B$.

\begin{figure*} [ht!]
	\begin{center}
		\includegraphics [width=1\textwidth]{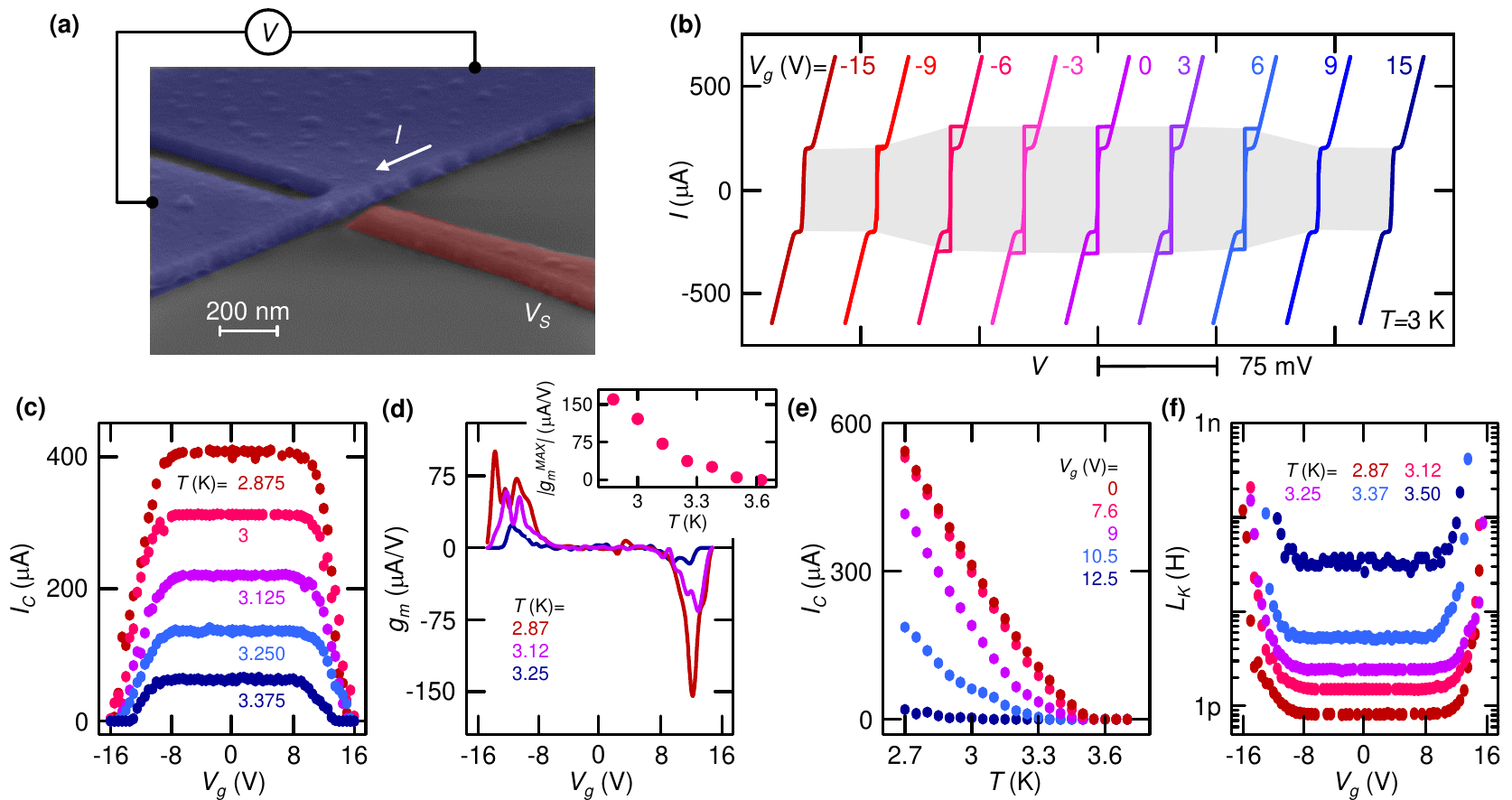}
	\end{center}
	\caption{\label{fig:Fig8}\textbf{Behavior of vanadium DB-FETs.} (a) False color scanning electron micrograph of a typical Ti-based DB-FET. The transistor channel (blue) is current biased and the voltage drop is measured in 4-wire configuration, while the gate voltage can be applied to the lateral gate electrode (red, $V_{S}$).(b) Current-voltage characteristics of a vanadium DB-FET measured at 3 K for several values of gate voltage ($V_g$). The curves are horizontally offset for clarity. The semitransparent gray area depicts the parameters space where superconductivity persists. (c) Dependence of critical current ($I_C$) on gate voltage ($V_g$) for different values of bath temperature ($T$). (d) Transconductance ($g_m$) as a function of gate voltage for different values of temperature ($T$). Inset: Absolute value of the maximum transconductance ($g_m^{MAX}$) as a function of temperature ($T$). (e) Dependence of critical current ($I_C$) on temperature ($T$) for different values of gate voltage ($V_g=$). (f) Josephson kinetic inductance ($L_K$) versus gate voltage ($V_g$) of a typical vanadium DB-FET for different values of temperature ($T$).}
\end{figure*}

\subsection{Vanadium DB-FETs}\label{DBV}
The study of vanadium-based DB-FETs encloses a two fold relevance. On the one hand, the demonstration of field-effect control of $I_C$ on JJs made of a different material points towards the universality of the phenomenon already presented for titanium. On the other hand, its relative high critical temperature ($T_C\sim5$ K \cite{Matthias1963}) elects vanadium as a strong candidate for industrial applications. 

Figure \ref{fig:Fig8}-a shows a false color scanning electron micrograph of a typical vanadium DB-FET. The nano-constriction is $\sim180$ nm-long, $\sim110$ nm-wide, and $\sim60$ nm-thick. The complete $I-V$ characteristics measured at a temperature $T= 3$ K for different values of gate voltage applied to the side gate electrode $V_g=V_S$ are shown in Fig. \ref{fig:Fig8}-b. The trace recorded at $V_g=0$ shows the usual hysteretic behavior due to the overheating related to the normal/superconducting state transition \cite{Courtois2008}. The difference between the switching and the retrapping currents is less pronounced than in the case of titanium Dayem bridges (see Fig. \ref{fig:Fig4}-b for comparison), because the higher measurement temperature ensures a more efficient cooling of the quasiparticles through phonons \cite{Giazotto2006,Courtois2008}. 
For both polarities of $V_g$, the critical current is monotonically suppressed by increasing the gate voltage. In full agreement with all the previously presented experiments, the normal state resistance of the vanadium DB-FETs is not affected by the applied electric field.

Figure \ref{fig:Fig8}-c shows the dependence of the critical current ($I_C$) on gate voltage ($V_g$) for different values of temperature. The $I_C$ vs $V_g$ characteristics show the typical features already recorded for titanium devices: a plateau of constant $I_C$ at low values of gate voltage followed by a steep decrease of the critical current down to its full suppression. For most of the investigated temperatures the critical gate voltage is $V_{g}^C\simeq15$ V, but at $T=3.375$ K it is reduced to $V_{g}^C\simeq13$ V. This behavior is different from what measured on titanium DB-FETs \cite{Paolucci2018} (see Fig. \ref{fig:Fig4}-c for comparison). Despite the latter difference, these experiments showed a similar phenomenology of titanium DB-FETs. As a consequence, we can conclude that this unconventional field-effect is not material related, but it may be intrinsic to any metallic BCS superconductor JJ (and wires).

Figure \ref{fig:Fig8}-d shows the $g_m$ versus $V_g$ traces for different values of bath temperature. The transconductance strongly depends on temperature. In particular, its maximum value ($g_m^{MAX}$) ranges from 150 $\mu$A/V at $T=2.875$ K to 0 at $T=3.6$ K (see the inset of Fig. \ref{fig:Fig8}-d). We would like to stress that the maximum values of transconductance extrapolated for vanadium devices exceeds by at least one order of magnitude those obtained for titanium ones (see Fig. \ref{fig:Fig6}-e for comparison) and semiconductor JJs \cite{Akazaki1996,Doh2005}. Therefore, vanadium might be a suitable candidate for industrial applications of the DB-FETs in superconducting digital electronics.

The temperature behavior of Josephson effect is resumed in Fig. \ref{fig:Fig8}-e, where the critical current monotonically suppresses by increasing the gate voltage. The limited range of explored temperatures does not allow a reliable fit of the data with Eq. \ref{eq:CritCurr}. Nevertheless, we can foresee a behavior similar to titanium DB-FETs: the transport in the JJ evolves from the Kulik-Omelyanchuk characteristic of a clean ballistic constriction \cite{Kulik1977} to the Ambegaokar-Baratoff temperature dependence of a tunnel junction \cite{Ambegaokar1963}. 

The changes in switching current of a Josephson junction can be remapped as modifications of its kinetic inductance ($L_K$). The dependence of $L_K$ on $V_g$ is shown in Fig. \ref{fig:Fig8}-f for different values of bath temperature. The kinetic inductance is modulated of more than two orders of magnitude by the gate bias. In particular, $L_K$ acquires a minimum value of about 800 fH (for $V_g=0$ and $T=2.875$ K) and a maximum value of $\sim400$ pH (at $V_g=14$ V and $T=3.37$ K). Therefore, the vanadium DB-FETs might be suitable, in principle, for the realization of metallic superconducting qubits \cite{Larsen2015,deLange2015}.

\section{Proximity Josephson junctions} \label{Proximity}

\begin{figure*} [ht!]
\begin{center}
\includegraphics [width=1\textwidth]{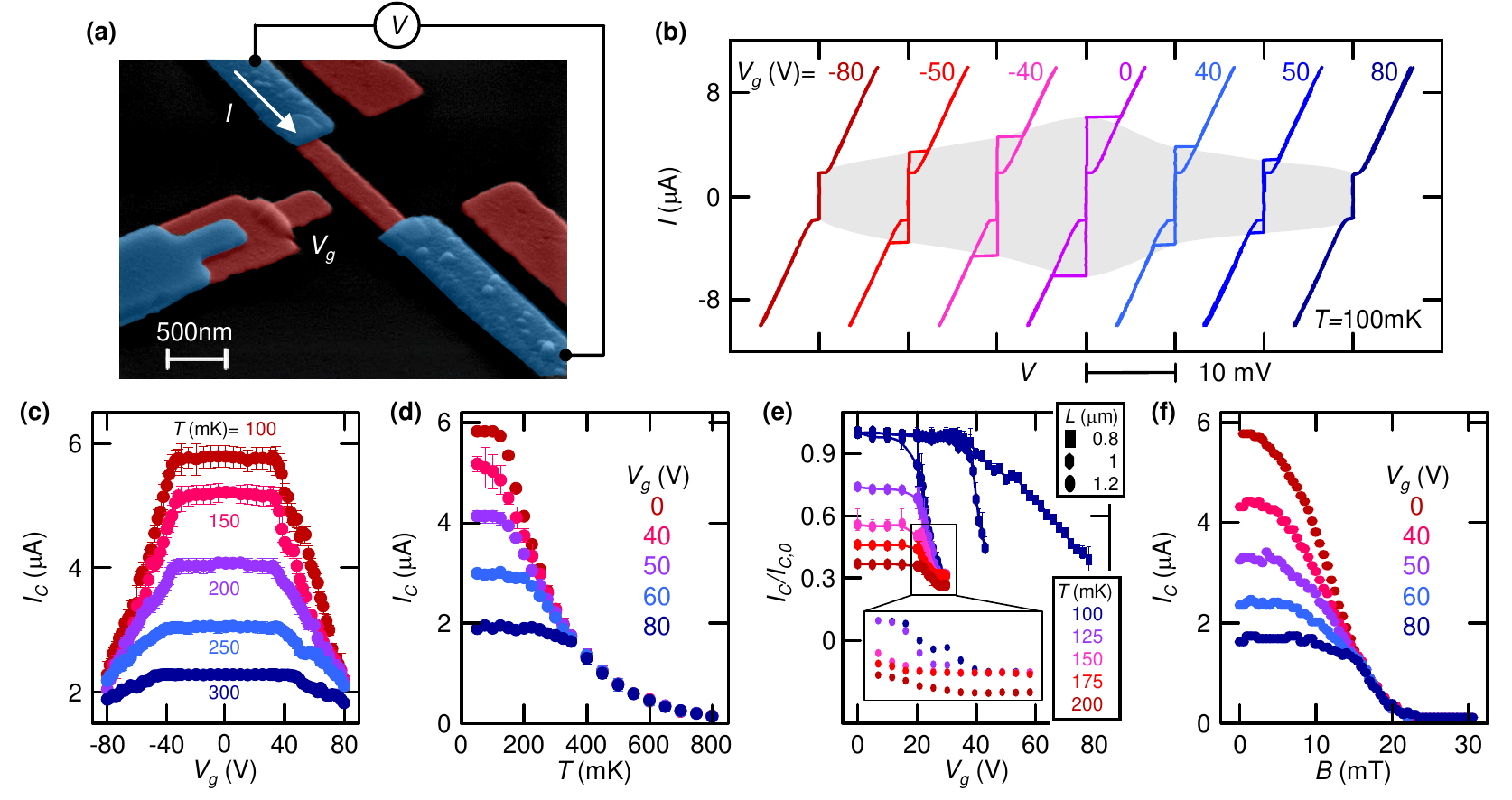}
\end{center}
\caption{\label{fig:Fig9}\textbf{Electric field dependence of supercurrent in SNS-FETs.} (a) False color scanning electron micrograph of a typical SNS-FET. The superconducting parent material is aluminum (blue), while the proximized normal metal is copper (red). The transistor is current biased ($I$) and the voltage drop ($V$) is measured in 4-wire configuration. The gate voltage ($V_g$) is applied through the side gate electrode.(b) Current-voltage characteristics ($I-V$) of a 800 nm long channel SNS-FET measured at 100 mK for several values of gate voltage ($V_g$). The curves are horizontally offset for clarity. The semitransparent gray area depicts the parameters space where superconductivity persists. (c) Dependence of critical current ($I_C$) on gate voltage ($V_g$) of a 800 nm long SNS-FET for different values of bath temperature ($T$). (d) Temperature dependence of critical current ($I_C$) of a 800 nm long SNS-FET measured for different values of gate voltage ($V_g=$). (e) Normalized critical current ($I_C/I_{C,0}$) as a function of gate voltage ($V_g$) of 800 nm long(rectangles), 1 $\mu$m long (hexagons) and 1.2 $\mu$m long (ellipses). For the latter a thermal characterization is shown. Inset: a blow-up of the saturation of $I_C$ with $V_g$. (f) Perpendicular-to-plane magnetic field dependence of the critical current ($I_C$) of a 800 nm long SNS-FET measured at $T=100$ mK.} 
\end{figure*}

This section presents a set of gating experiments realized on fully metallic superconductor-normal metal-superconductor Josephson junction-based field-effect transistors (SNS-FETs) \cite{DeSimoni2019}. We show the influence of temperature and magnetic field on this gating effect measured on SNS-FETs of different length. These experiments demonstrate that this unconventional field-effect does not require the presence of a proper pairing potential. On the contrary, the existence of induced superconducting correlations seems to be enough to enable the electric field control of supercurrent in the system.

The SNS-FETs typically consist of a Al/Cu/Al planar Josephson junction with the possibility of applying a gate bias ($V_g$) through a side gate electrode (see Fig. \ref{fig:Fig9}-a). Devices with 3 different inter-S-electrodes spacing ($L$), i.e. the length of the copper weak link, have been realized. In particular, data obtained for $L_A=800$ nm (type-$A$), $L_B=1$ $\mu$m (type-$B$) and $L_C=1.2$ $\mu$m (type-$C$) have been showed \cite{DeSimoni2019}.

The characteristic energy scale for the electronic diffusion in a SNS Josephson junction is the Thouless energy, which is defined as $E_{Th}=\hbar D/L^2$, where $\hbar$ is the reduced Planck constant and $D$ is the diffusion coefficient of the normal metal (for copper thin films $D\simeq0.008$ m$^2/$s \cite{DeSimoni2019}). The Thouless energy associated to the studied samples was $E_{T,A}\simeq 6.6$ $\mu$eV, $E_{T,B}\simeq 4.2$ $\mu$eV and $E_{T,A}\simeq 2.9$ $\mu$eV. Since for all of them $E_{Th} \ll \Delta$ (where $\Delta\simeq 180$ $\mu$eV is the superconducting energy gap of aluminum), all the weak links operate in the long-junction limit. 

Figure \ref{fig:Fig9}-b shows the $I-V$ characteristics of a typical type-$A$ SNS-FET measured at $T=100$ mK for several values of gate voltage ($V_g$). For $V_g=0$ the current versus voltage characteristic highlights the presence of Josephson effect with critical current $I_C\simeq 5.8$ $\mu$A and retrapping current $I_R\simeq1.9$ $\mu$A. In analogy with the phenomenology presented for genuine superconductors \cite{DeSimoni2018,Paolucci2018,Paolucci2019}, a clear suppression of $I_C$ has been recorded at high values of gate voltage ($V_g\geq40$ V), while the normal state resistance is completely unaffected by the external electric field. Furthermore, the decrease of critical current is symmetric in the polarity of the gate bias. As a consequence, charge accumulation/depletion can not account for the effect.

The complete temperature dependence of field-effect recorded for the same type-$A$ SNS-FET is resumed in Fig. \ref{fig:Fig9}-c. Differently from proper superconductors \cite{DeSimoni2018,Paolucci2018,Paolucci2019}, at fixed temperature the critical current monotonically decreases with increasing $V_g$ without reaching its full suppression in the studied gate voltage range. At the same time, the plateau of $I_C$ with $V_g$ widens with rising temperature.
 
Field-effect becomes completely inefficacious for $T\geq350$ mK (see Fig. \ref{fig:Fig9}-d), that is less than one third of the aluminum superconducting critical temperature. In fact, at low temperature the $I_C(V_g)$ strongly deviates from the unperturbed case, while all the curves are overlapped for $T\geq350$ mK. We also notice that the electric field generated by the gate electrode has been calculated to impact exclusively on the copper wire without reaching the Al leads \cite{DeSimoni2019}. 
These results seem to suggest that the field-effect does not require a proper genuine superconductor but it can occurs even in the presence of superconducting correlations.

The dependence of the field-effect on the length of the normal metal strip is summarized in Fig. \ref{fig:Fig9}-e, where the normalized critical current ($I_C/I_{C,0}$) is plotted as a function of the gate voltage for representative type-$A$, type-$B$ and type-$C$ devices. On the one hand, the plateau of constant critical current shortens with increasing $L$. On the other hand, a saturation region for high values of gate voltage appears for the type$-C$ devices in the complete temperature range studied (see the inset of Fig. \ref{fig:Fig9}-e). This behavior is strictly peculiar of SNS-FETs, since gate-induced full suppression of the critical current has been demonstrated both in metallic superconductor wires \cite{DeSimoni2018} and DB-FETs \cite{Paolucci2018, Paolucci2019} (see Sections \ref{Wires} and \ref{Josephson}, respectively). As a consequence, these experiments suggest a nontrivial relation between the Thouless energy of the JJ and the ability of the electric field to affect its critical current. In addition, the saturation of $I_C$ for high values of gate voltage suggests once more the exclusion of quasiparticles overheating as leading mechanism of the measured effect \cite{Morpurgo1998, Savin2004, Roddaro2011}. 

The combined action of an electric and magnetic field on the critical current of the SNS-FETs has been investigated by measuring $I_C$ as a function of the perpendicular-to-plane magnetic field ($B$) for different values of bias applied to the lateral gate electrode ($V_g$), as depicted in Fig. \ref{fig:Fig9}-f. 
Similarly to the thermal characterization (see Fig. \ref{fig:Fig9}-d), for $B \geq 15$ mT the curves at different values of gate voltage overlap. In other words, the impact of this unconventional field-effect on the switching current saturates at high values of magnetic field. These observations confirm a damping of the electrostatic effect on the weak link as it approaches the normal state. Further details can be found in Ref. \cite{DeSimoni2019}.

These results obtained on SNS-FETs suggest that the impact of the electric field is initially more relevant in weaker superconductors, but as the system approaches the resistive state (by temperature or magnetic field increases) a threshold is reached above which no further suppression of $I_C$ can be observed (the field-effect becomes unefficient). 

\begin{figure*} [ht!]
	\begin{center}
		\includegraphics [width=1\textwidth]{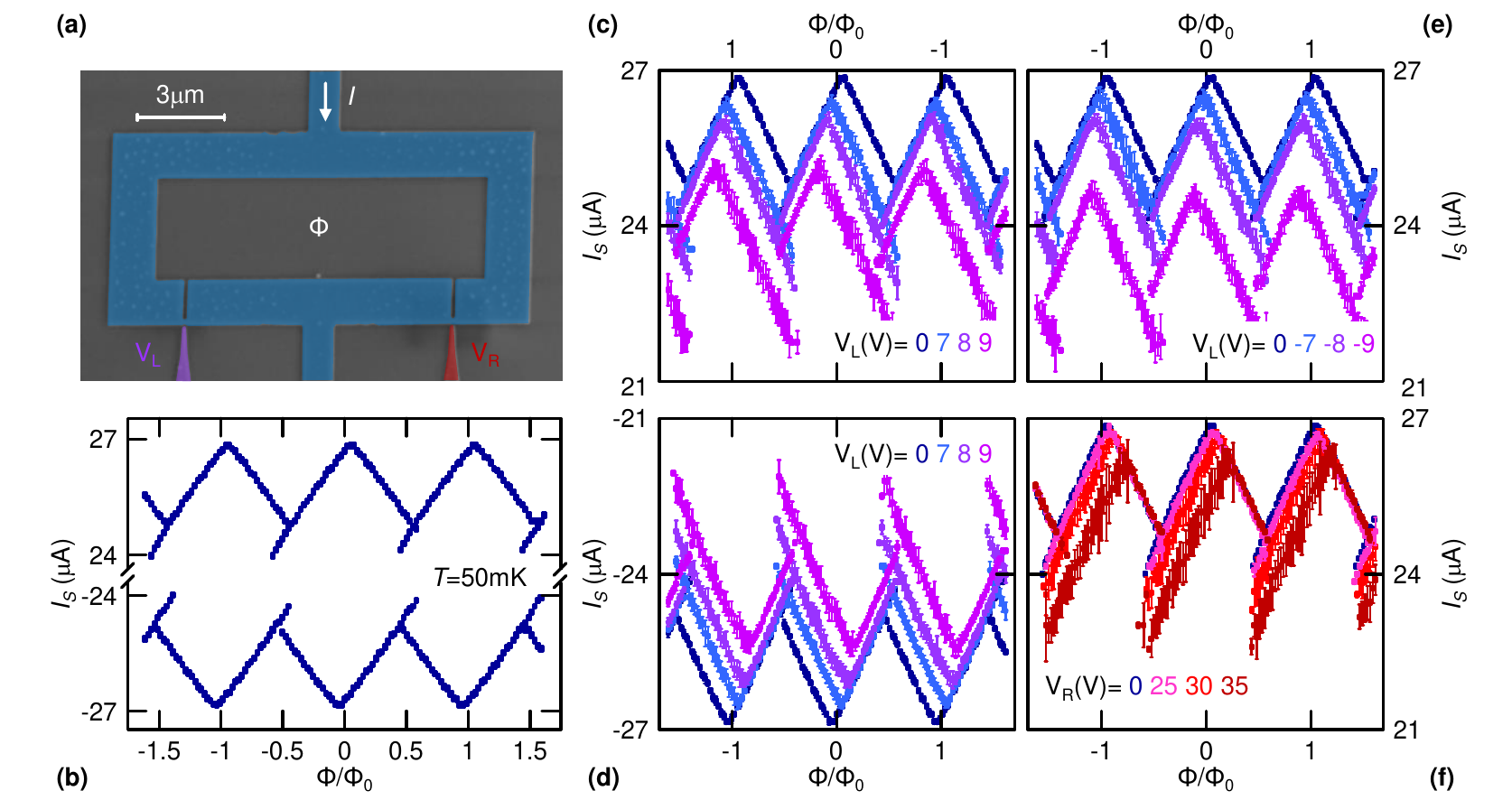}
	\end{center}
	\caption{\label{fig:Fig10}\textbf{Electric field control of the interference pattern of a metallic SQUID.} (a) False-color scanning electron micrograph of a typical field-effect controllable Josephson interferometer. The core of the SQUID is represented in blue whereas the left ($V_L$) and right ($V_R$) side gate electrodes are shown in purple and red, respectively. $I$ is the total current flowing through the interferometer, while $\Phi$ represents the magnetic flux piercing the loop. (b) Switching current ($I_S$) of the interferometer as a function of the normalized flux piercing the loop ($\Phi/\Phi_0$) measured at $T=50$ mK. (c) Modulation of the switching current ($I_S$) with the magnetic flux $\Phi$ piercing the SQUID at $T=50$ mK for positive current bias measured for positive voltages applied to the left gate electrode ($V_L$). (d) Negative branch of $I_S(\Phi)$ measured for positive values of $V_L$. (e) Positive branch of $I_S(\Phi)$ for different negative values of $V_L$ and for different values of positive right gate bias $V_R$ (f) measured at $T=50$ mK.} 
\end{figure*}

\section{SQUID} \label{SQUID}

Here, we resume transport experiments performed on fully metallic SQUIDs where each JJ was independently mastered through a voltage applied to side gate electrodes placed at each Dayem bridge \cite{Paolucci2019_2}. The electric field has been shown to control the whole interference pattern of the SQUID. In particular, the switching current ($I_S$) was suppressed to values lower than the critical current of a single JJ. This has been explained assuming that the electric field \textit{couples} (even if not directly) with the superconducting phase.

A typical metallic gate-tunable Josephson interferometer is composed by a titanium superconducting loop (30 nm thick) interrupted by two nanoconstrictions (150 nm long and 150 nm wide) separated of about 8 $\mu$m, as shown in Fig. \ref{fig:Fig10}-a. At each JJ, a side gate electrode ($V_L$ and $V_R$ for the left and right junction, respectively) allows the application of an electrostatic field in the Dayem bridge region. This enables the independent control of the switching current of the two Josephson junctions ($I_L$ and $I_R$), as confirmed by finite element simulations of the electric field distribution for the same geometry of the device \cite{Paolucci2019_2}.

The current-flux behavior of the SQUID showed the characteristic triangular patterns typical of a Dayem bridge-based interferometer \cite{Fulton1972, Tsang1975} for both the positive and negative branch of the switching current, as represented in Fig. \ref{fig:Fig10}-b. The fit of the experimental data with a theoretical model based on the resistively and capacitively shunted junction (RCSJ) formalism allows us to obtain the SQUID basic parameters. In particular, the critical current of the two junctions was shown to be slightly asymmetric ($I_{C,L}\simeq 13.6$ $\mu$A and $I_{C,L}\simeq 13.3$ $\mu$A), and the screening parameter was calculated to reach the noticeable value $\beta \simeq32$ since the kintic inductance \cite{Annunziata2010} of the Ti ring was as high as $\mathcal{L}_K\simeq0.6$ nH.

\begin{figure} [t!]
	\begin{center}
		\includegraphics [width=1\columnwidth]{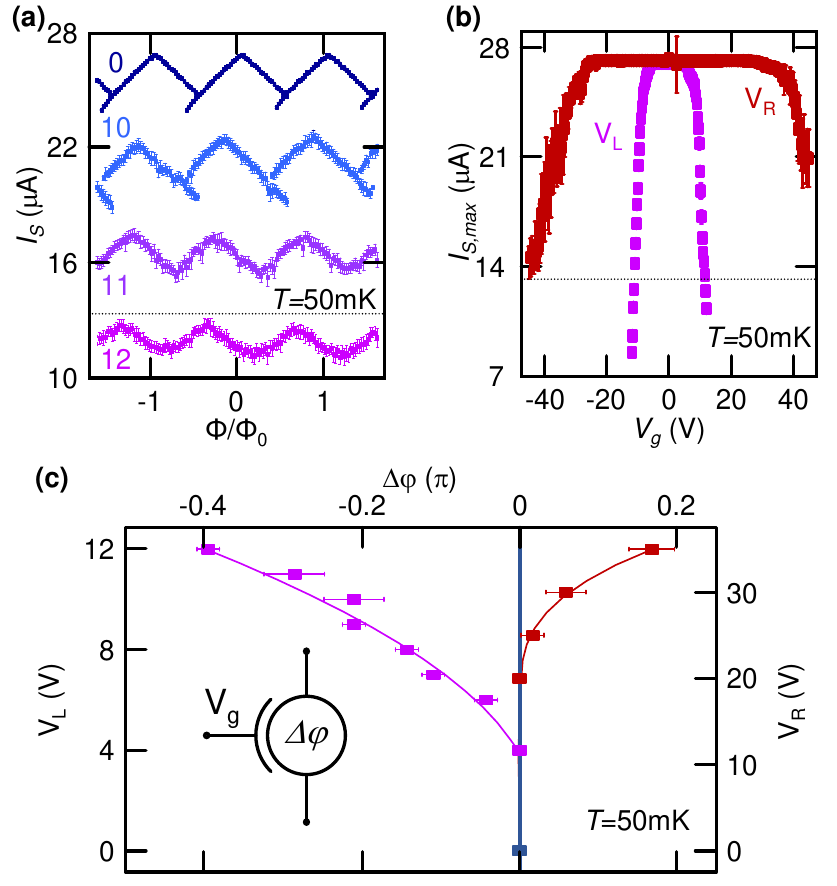}
	\end{center}
	\caption{\label{fig:Fig11}\textbf{High gate voltage regime and tunable phase shift.} (a)  Modulation of the switching current ($I_S$) with the magnetic flux $\Phi$ piercing the SQUID at $T=50$ mK for positive voltages applied to the left gate electrode ($V_L$). The  horizontal black line depicts the critical current of the right JJ. (b)  Switching current at $\Phi=0$ versus the gate voltage applied to the left (violet squares) and right (red squares) electrode at $T = 50$ mK. (c) Phase shift versus the left  (violet squares) and right (red squares) gate electrode voltage extracted from data at T = 50 mK. Inset: schematic representation of a gate-tunable phase shifter.}
\end{figure}

The gate bias has a strong impact on the $I_S(\Phi)$ characteristic of the SQUID, as shown in Fig. \ref{fig:Fig10}-c,d for positive values of $V_L$ while the right gate electrode was left grounded or floating. The data highlight the suppression of the maximum switching current [$I_{S,max}=I_S(\Phi=0)$] for both current biases and a shift of $I_{S,max}$ along the magnetic flux axis in opposite direction for the two switching current branches. Furthermore, the error bars of $I_S$ are different in the two halves of the interference pattern period.

In accordance with results obtained on superconducting wires \cite{DeSimoni2018} (see Section \ref{Wires}), Dayem bridges \cite{Paolucci2018, Paolucci2019} (see Section \ref{Josephson}) and proximity JJs \cite{DeSimoni2019} (see Section \ref{Proximity}), the switching current suppression is almost symmetric in the polarity of the applied gate bias (see Fig. \ref{fig:Fig10}-e). 

A gate voltage applied to the right gate electrode ($V_R$) generates a decrease of $I_{S,max}$ (as for $V_L$), shifts the interference pattern in the opposite direction along the magnetic flux axis than for the left gate electrode, and increases the current fluctuations in the second half period of $I_S(\Phi)$ (see Fig. \ref{fig:Fig10}-f). The different values of $V_L$ and $V_R$ necessary to affect $I_S$ can be attributed to different distance between the JJ and the gate electrode in $L$ and $R$ \cite{Paolucci2019_2}.

The conventional models for SQUIDs predict a flux-independent current flow when the critical current of one of the two junctions is suppressed down to full closure \cite{Clarke2004,Barone1982}. By contrast, $I_S(\Phi)$ has been shown to be lowered by the application of $V_L=12$ V below the critical current of a single Josephson junction, while preserving its oscillatory behavior \cite{Paolucci2019_2} (see Fig. \ref{fig:Fig11}-a).

Figure \ref{fig:Fig11}-b shows the gate voltage dependence of $I_{S,max}$. The plot highlights once more that the total SQUID switching current reduces under the critical current of a single JJ. In addition, we notice that the electric field induced switching current suppression seems to be more efficient for negative values of gate voltage in both Dayem bridges. In particular, the zero-flux value of the switching current reaches $I_{S,max}\simeq0.6I_{C,R}$ for $V_L=-12$ V. 

The phenomenology resumed in Fig. \ref{fig:Fig11}-a,b indicates that an electrostatic field applied to a single Dayem bridge indirectly affects the other junction and, therefore, the entire interferometer. This behavior can be obtained with a model that imposes gate-dependent suppression of the critical current and creation of phase fluctuations of a single JJ \cite{Paolucci2019_2}. Since the phases of the two JJs are locked by fluxoid quantization, these local perturbations can affect the phase of the other junction and, as a consequence, the entire interferometer. This model suggests that an external electrostatic field \textit{couples} (even if indirectly) with the superconducting phase of a conventional metallic superconductor.

The gate-tunable interferometer can be used as a controllable phase shifter (see the inset of Fig. \ref{fig:Fig11}-c) \cite{Paolucci2019_2}, where the phase shift is defined as $\Delta\varphi(V_i)=\varphi(V_i)-\varphi(0)$ with $V_i$ the gate bias applied to the $i=L,R$ electrode. Figure \ref{fig:Fig11}-c shows $\Delta\varphi$ generated by the application of $V_L$ (violet) and $V_R$ (red). The effect on the phase shift is asymmetric, because the two gate electrodes are placed at different distances from the corresponding JJ. In fact, we notice $\Delta\varphi_{max}(V_L)\simeq-0.4$ while $\Delta\varphi_{max}(V_R)\simeq0.2$. 
The maximum achievable phase shift may be improved by applying higher values of gate voltage, under the condition of preserved oscillating $I_S(\Phi)$. 

Concluding, the experiments on the gate-tunable metallic Josephson interferometers show the \textit{coupling} between an external electric field and the macroscopic superconducting phase. In addition, this device is a promising candidate for the realization of gate-tunable flux \cite{Mooij1999, Yan2016} and phase \cite{Martinis2002} superconducting qubits \cite{Koch2007, Schreier2008, Barends2013}, and for the implementation of superconducting electronics \cite{Terzioglu1998} in the form of rapid single flux quantum (RSFQ) logic \cite{Likharev1995,Worsham1998}.

\section{Possible applications} \label{Applications}
There are several possible applications arising from the phenomenology related to the field-effect control of the supercurrent in all the physical systems presented along this review.
These works may open the way for a number of novel gate-tunable devices, such as magnetometers \cite{Clarke2004,Giazotto2010,Strambini2016}, heat mastering systems \cite{Giazotto2012,Mart2014, Perez2014, Fornieri2015,Fornieri2017}, radiation detectors \cite{Gol2001}, and metallic architectures for quantum \cite{Larsen2015,deLange2015, Mooij1999,Yan2016,Martinis2002} and classical \cite{Walker1963,McCaughan2014, Zhao2017} computation.
In particular, Sec. \ref{Qubits} will show possible implementations of superconducting qubits, while Sec. \ref{Electronics} is devoted to classical electronics which exploits this unconventional gating effect.

\begin{figure} [t]
	\begin{center}
		\includegraphics [width=1\columnwidth]{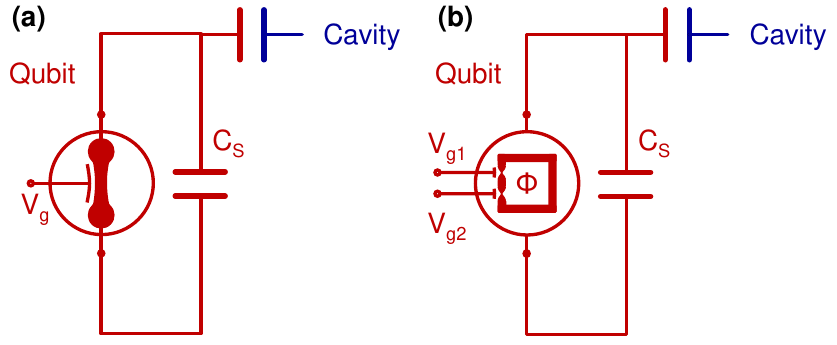}
	\end{center}
	\caption{\label{fig:Fig12}\textbf{Gate-tunable metallic superconductor qubits.} The Dayem bridge transmon \textit{dayemon} (a) and the SQUID transmon \textit{squidmon} (b), where the qubit is capacitively coupled to a resonant cavity. The \textit{dayemon} can be controlloed by the gate voltage ($V_g$). The \textit{squidmon} can be independently mastered through the magnetic flux ($\Phi$) and two gate voltages ($V_{g1}$ and $V_{g2}$).}
\end{figure}

\subsection{Qubits} \label{Qubits}

\begin{figure*} [ht!]
	\begin{center}
		\includegraphics [width=1\textwidth]{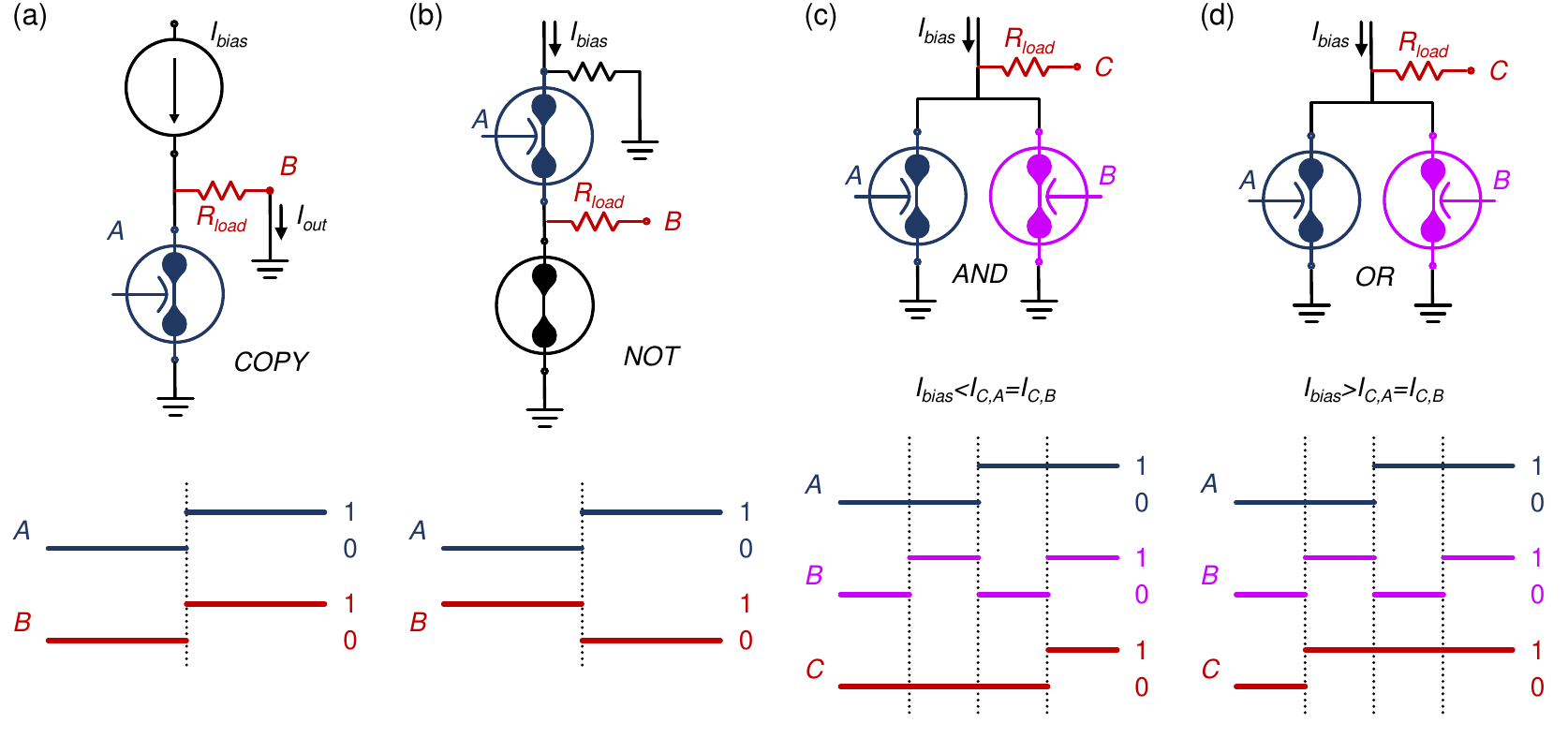}
	\end{center}
	\caption{\label{fig:Fig13}\textbf{Gate-tunable metallic superconducting electronics.} Basic logic gates employing the electric-field-controlled cryotron \textit{EF-tron}: COPY (a), NOT (b), AND (c), and OR (d). The truth table of each individual circuit is represented.}
\end{figure*}

The transmission line shunted plasma oscillation qubit (transmon) \cite{Koch2007,Wendin2017} is the leading architecture for the implementation of superconducting qubits. It consists of a Cooper pair box (where one more JJs are shunted by a capacitance $C_S$) capacitively coupled to a resonant cavity (usually a $LC$ circuit) that is driven by microwave radiation. 
The interesting energy scales are the Josephson coupling energy $E_D=\hbar I_C/2e$ \cite{Barone1982} and the charging energy $E_C= 2 e^2/C_{total}$, where $C_{total}$ is the total capacitance.
In order to minimize the charge noise affecting the qubit, the value of $C_S$ (and therefore $C_{total}$) needs to be chosen in order to fulfill $E_D\gg E_C$, i.e. the Josephson coupling energy has to be much greater than the charging energy. At the same time, the Josephson energy can be tuned by means of an external magnetic flux threading a SQUID \cite{Koch2007} or by a gate bias coupled to a proximized semiconductor nanowire \cite{Larsen2015,deLange2015}.

The tunability over two orders of magnitude of the Josephson critical current in metallic Dayem bridges (see Sec. \ref{Josephson}) is reflected in the possibility of controlling $E_D$ with an electric field. As a consequence, it is possible to envision a fully metallic gate-tunable Dayem bridge transmon: the \textit{dayemon} (see Fig. \ref{fig:Fig12}-a). 

A more exotic and versatile architecture might employ the gate-tunable SQUID (presented in Sec. \ref{SQUID}) to control the Josephson coupling: the \textit{squidmon}, as depicted in Fig.  \ref{fig:Fig12}-b. This device would increase the number of knobs able to control the qubit state. In fact, the switching current of this type of interferometer can be independently tuned both with the magnetic flux and two gate electrodes. 

The performances of \textit{dayemons} and \textit{squidmons} are predicted to be similar to that of conventional transmon qubits \cite{Paolucci2019} if the gating mechanism does not introduce large noise (that would causes decoherence of the qubit). Furthermore, their \textit{monolithic} architecture (single step fabrication process without heterojunctions between different materials) might promote them as a promising scalable technology for the superconducting qubit implementation.

\subsection{Superconducting classical electronics} \label{Electronics}

Superconducting classical computation hinges upon the modulation of the critical current of a transistor through an external gate bias \cite{Likharev2012}. The challenge has been addressed by different approaches: the superconducting field-effect transistor (SuFET) \cite{Nishino1989,Fiory1990,Mannhart1993,Okamoto1992,Mannhart1993b}, where the critical current of a low charge carrier density superconducting thin film is modulated by a gate voltage; the Josephson field-effect transistor (JoFET) \cite{Clark1980,Takayanagi1985,Akazaki1996}, where $I_C$ of a proximized semiconductor thin film is controlled via field-effect; and the nanocryotrons (nTrons) \cite{Walker1963,McCaughan2014,Zhao2017}, where a gate current tunes the supercurrent transport in a metallic channel.

The electric field control of supercurrent in metallic superconductor wires (see Sec. \ref{Wires}) could be used to realize an electric-field-controlled cryotron (\textit{EF-tron}), as depicted in Fig. \ref{fig:Fig13}-a for a COPY logic circuit. 
The device is current biased under the wire pristine critical current ($I_{bias}<I_{C0}$) and the normal state resistance of the wire is much higher than the load ($R_N\gg R_{load}$). For $V_A=0$ (logic state 0) all the current flows in the transistor channel (the output current is zero, $I_{out}=0$) and the output voltage is 0 (logic state 0). On the contrary, for $I_{bias}>I_C(V_A)$ (input logic state 1) the \textit{EF-tron} switches in the normal state, the output current grows ($I_{out}\neq 0$) and the output shows logic state 1.

Similarly, the NOT logic gate can be realized by employing the \textit{EF-tron}, as shown in Fig. \ref{fig:Fig13}-b. It is composed by the series of a gate-tunable wire (controlled by the input A) and a constriction of critical current lower than $I_{bias}$, and a load resistance $R_{load} \ll R_N$ (with $R_N$ normal state resistance of the ungated constriction).
When the input $A$ is $0$ the current can pass through the first wire, while the second constriction switches to the normal state. As a consequence, the current flows through $R_{load}$ and the output signal is $1$. On the contrary, when $A=1$ the wire is in the normal state. Therefore, the current does not flow in the device core and the output signal is $0$.

The \textit{EF-tron} could be used to define a functionally complete logic, since it can realize the conjunction (AND, see Fig. \ref{fig:Fig13}-c) and disjunction (see Fig. \ref{fig:Fig13}-d) gate. Both devices are realized by the parallel connection of two electric-field controlled cryotrons characterized by the same intrinsic critical current ($I_C=I_{C,A}=I_{C,B}$). 
The structure provides the AND logic gate when the bias is lower than the pristine critical current of a single \textit{EF-tron} ($I_{bias}<I_C$). In fact, the current flows to the output electrode (i.e C=1) only when both inputs are at logic state 1 (both \textit{EF-tron}s are in the normal state). 
By contrast, when $I_{bias}>I_C$ a single input set to logic state 1 causes the transition of both \textit{EF-tron} channels to the normal state. Therefore, the device exhibits the OR behavior.

\begin{figure} [t]
	\begin{center}
		\includegraphics [width=1\columnwidth]{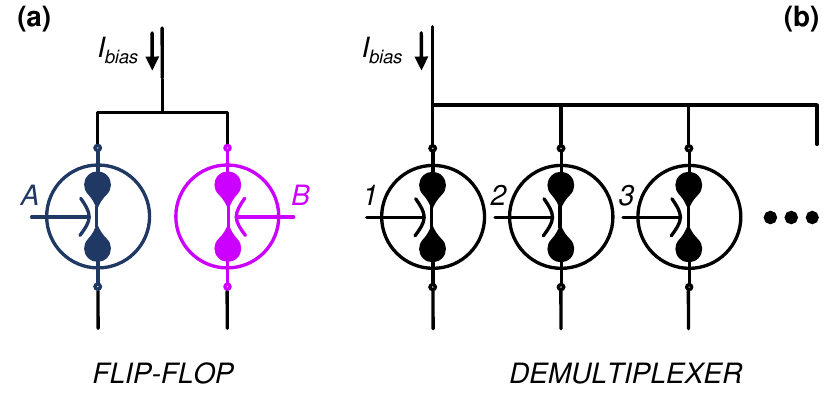}
	\end{center}
	\caption{\label{fig:Fig14}\textbf{More complex electronic circuits based on the \textit{EF-tron}.} Schematic representation of a flip-flop (a) and a demultiplexer (b) circuits.}
\end{figure}

The \textit{EF-tron} could be also used to implement more complex functions and circuits. For instance, the parallel connection of two cryotrons could implement the flip-flop storage circuit (see Fig. \ref{fig:Fig14}-a). The bias current ($I_{bias}$) can be directed in the $A$ (or $B$) branch according to the logic state given to the two \textit{EF-tron}s. Interestingly, the directed current continues to flow through the $A$ ($B$) branch even if the write pulse to $B$ ($A$) is removed, since the absence of dissipation prevents current redistribution \cite{Walker1963}. As a consequence, this circuit could serve as information storage loop in a fully metallic superconducting electronics architecture.

A similar approach leads to the creation of a demultiplexer, where the current injected from a single input electrode can be directed to one of the several outputs, as depicted in Fig. \ref{fig:Fig14}-b. In particular, the role of each electric-field-controlled cryotron is to forbid (allow) the supercurrent flow through the corresponding branch. 

Concluding, the \textit{EF-tron}s marry the easy single step fabrication process of a nanocryotron \cite{Walker1963,McCaughan2014,Zhao2017} with the very large input-to-output impedance ($\sim 1-10$ T$\Omega$) of a SuFET \cite{Nishino1989,Fiory1990,Mannhart1993,Okamoto1992,Mannhart1993b} (or a JoFET \cite{Clark1980,Takayanagi1985,Akazaki1996}). Therefore, they could be the cornerstone of a new and scalable superconducting electronics.

\section{Summary and future directions}\label{Concl}

Along this review we showed the electric field control of the critical current in different systems made of conventional metallic superconductors: titanium and aluminum wires (Section \ref{Wires}), titanium and vanadium nano-constriction Josephson junctions (Section \ref{Josephson}), aluminum/copper/aluminum proximity JJs (Section \ref{Proximity}), and titanium superconducting quantum interference devices based on nano-constrictions (Section \ref{SQUID}). Despite a microscopic explanation of this effect is not available yet, the presented experiments provide a wide platform of discussion and analysis. 

The most prominent common feature in all the studied systems is the suppression of critical supercurrent symmetric in the polarity of gate voltage. In addition, the normal state resistance of the various systems is not affected by the electric field. As a consequence, the electric field does not act on the carrier concentration at the materials surface, \textit{i.e.} charge accumulation/depletion can be excluded.

On the one hand, at low temperature the experiments highlighted a propagation of the effect into the superconductor for several times the superconducting coherence length, which decreases while approaching the critical temperature. On the other hand, the effect of two gate electrodes placed at the opposite sides of the superconductor does not sum up. Therefore, the electric field seems to stop at the superconductor surface, but the supercurrent is non-locally affected deep in its bulk thank to the coherence of the superconducting condensate.

Hot quasiparticles injection in superconducting nano-structures is an ordinary source of critical current reduction, since it increases their effective electron temperature and, consequently, damps the superconducting energy gap. Experiments performed on samples fabricated onto different insulating substrates highlighted the complete uncorrelation between the $I_C$ reduction and the leakage current. In addition, the critical temperature of DB-FETs is completely gate voltage independent, and SNS-FETs with long normal metal wire showed saturation at high values of gate voltage (and leakage current). Therefore, simple quasiparticle overheating seems to be excluded.

The partial suppression of the switching current in proximity Josephson junctions demonstrates that the presence of a proper pairing potential is not required, but the existence of induced superconducting correlations enables this unconventional field-effect. Furthermore, the saturation of the effect seems to indicate an increase of the resilience of the Josephson coupling to the applied electric field when the weak link approaches the normal state.

The gating experiments performed on the SQUIDs cannot be explained by simple squeezing of the critical current of a single junction. In particular, the suppression of the interferometer switching current to values lower than the critical current of a single JJ can be only explained assuming that the an electrostatic field \textit{couples} with the phase dynamics of a superconductor.

To shed light on these gating experiments and move towards the understanding of the intimate relation between superconductivity and electric field, a set of complementary experiments would be necessary. For example, scanning tunneling microscopy or scanning gate experiments could provide information about the spatial variation of the superconducting gap due to presence of an external electric field. In addition, scanning SQUID microscopy would render the gate-dependent supercurrent spatial distribution. 

Instead, the electric field-driven transition mechanism could be studied by investigating the distribution function of the switching current, that is the escaping current from the superconducting to the normal state. Furthermore, the change in critical current, the phase dymamics and possible quasiparticle poisoning due to the gating effect could be studied through radio-frequency-based experiments, such as measurements of the resonance frequency of a gated circuit, and of Shapiro steps in Josephson junctions. 

In parallel, the puzzle needs to be tackled from a theoretical point of view. Despite a few phenomenological models have been proposed \cite{DeSimoni2018,Paolucci2019,Paolucci2019_2}, a microscopic picture is still missing. The first attempt to provide a microscopic explanation proposes a mechanism based on the electric field-driven modification of the orbital Rashba coupling at the surface \cite{Mercaldo2019}. Besides this, the knowledge of the physics at the origin of this unconventional gating is quite limited. 

Even without a theoretical explanation, the field-effect control of supercurrent in metallic BCS materials could be employed to develop devices with a plethora of applications. In particular, in Section \ref{Applications} we showed a possible vision of fully metallic classical and quantum computation architectures. The first goal of applied research lines might be to demonstrate the feasibility of such systems by realizing the first proof of principle devices, such as \textit{gatemons} or \textit{EF-trons}. Successively, large-scale superconducting electronics systems working in sinergy with conventional computing technologies could be designed by employing higher critical temperature superconductors (such as niobium and niobium nitride) compatible with present industrial standards.

\begin{acknowledgments}
We acknowledge  A. Braggio, P. Virtanen and R. Cristiano for fruitful discussions.
The authors acknowledge the European Research Council under the European Unions Seventh Framework Programme (FP7/2007-2013)/ERC Grant No. 615187 - COMANCHE, the European Unions Horizon 2020 research and innovation programme under the grant No. 777222 ATTRACT (Project T-CONVERSE) and Horizon 2020 and innovation programme under grant agreement No. 800923-SUPERTED for partial financial support. The work of F.P. work has been partially supported by the Tuscany Government (POR FSE 2014 -2020) through the INFN-RT2 172800 Project. 
\end{acknowledgments}


\begin{thebibliography}{99}

\bibitem{London1935}
F. London, and H. London,
Proc. R. Soc. London A {\bf{149}}, 71 {(1935)}.

\bibitem{London1936}
H. London,
Proc. R. Soc. London A {\bf{155}}, 102 {(1936)}.

\bibitem{vonLaue1935}
M. von Laue, F. London, and H. London,
Z. Phys. {\bf{96}}, 359 {(1935)}.

\bibitem{London1937}
F. London,
\textit{Superfluids},vol. I, (Wiley, New York, 1950).

\bibitem{Tao2002}
R. Tao, X. Xu, Y. C. Lan, and Y. Shiroyanagi, 
Physics C {\bf{377}}, 357 {(2002)}.

\bibitem{Moro2003}
R. Moro, X. Xu, S. Yin, and W. A. de Heer,
Science {\bf{300}}, 1265 {(2003)}.

\bibitem{Larkin1963}
A. I. Larkin, and A. B. Midgal,
Sov. Phys. JEPT {\bf{17}}, 1146 {(1963)}.

\bibitem{Lang1970}
N. D. Lang, and W. Kohn,
Phys. Rev. B {\bf{1}}, 4555 {(1970)}.

\bibitem{Ummarino2017}
G. A. Ummarino, E. Piatti, D. Daghero, R. S. Gonnelli, I. Y. Sklyadneva, E. V. Chulkov, and R. Heid,
Phys. Rev. B {\bf{96}}, 064509 {(2017)}.

\bibitem{Virtanen2019}
P. Virtanen, A. Braggio, and F. Giazotto,
arXiv:1903.01155  {(2019)}

\bibitem{Shapiro1984}
B. Y. Shapiro,
Phys. Lett. A {\bf{105}}, 374 {(1984)}.

\bibitem{Burlachkov1993}
L. Burlachkov, I. B. Khalfin, and B. Y. Shapiro,
Phys. Rev. B {\bf{48}}, 1156 {(1993)}.

\bibitem{Lee1996}
W. D. Lee, J. L. Chen, T. J. Yang, and B.-S. Chiou,
Physica C {\bf{261}}, 167 {(1996)}.

\bibitem{Lipavsky2006}
P. Lipavsk\'y, K. Morawetz, J. Kol\'a\v cek, and T. J. Yang,
Phys. Rev. B {\bf{73}}, 052505 {(2006)}.

\bibitem{Morawetz2008}
K. Morawetz, P. Lipavsk\'y, J. Kol\'a\v cek, and E. H. Brandt,
Phys. Rev. B {\bf{78}}, 054525 {(2008)}.

\bibitem{Bonfiglioli1956}
G. Bonfiglioli, and R. Malvano,
Phys. Rev. {\bf{101}}, 1281 {(1956)}.

\bibitem{Bonfiglioli1959}
G. Bonfiglioli, and R. Malvano,
Phys. Rev. {\bf{115}}, 330 {(1959)}.

\bibitem{Berman1975}
A. Berman, and H.~J. Juretschke,
Phys. Rev. B {\bf{11}}, 2903 {(1975)}.

\bibitem{Glover1960}
R. Glover, and M. Sherrill,
Phys. Rev. Lett. {\bf{5}}, 248 {(1960)}.

\bibitem{Bonfiglioli1962}
G. Bonfiglioli, R. Malvano, and B.~B. Goodman,
J. Appl. Phys. {\bf{33}}, 2564 {(1962)}.

\bibitem{Locklin2003}
J. Locklin, K. Shinbo, K. Onishi, F. Kaneko, Z. Bao, and R. C. Advincula,
Chem. Mater. {\bf{15}}, 1404 {(2003)}.

\bibitem{Panzer2005}
M.~J. Panzer, C.~R. Newman, and C.~D. Frisbie, 
Appl. Phys. Lett. {\bf{86}}, 103503 {(2005)}.

\bibitem{Kirby2013}
B. J. Kirby,
\textit{Micro- and Nanoscale Fluid Mechanics}, (Cambridge University Press, Cambridge, 2013).

\bibitem{Dhoot2006}
A.~S. Dhoot, J.~D. Yuen, M. Heeney, I. McCulloch, D. Moses, and  A.~J. Heeger,
Proc. Natl. Acad. Sci. U.S.A. {\bf{103}}, 11834 {(2006)}.

\bibitem{Misra2007}
R. Misra, M. McCarthy, and A. F. Hebard,
Appl. Phys. Lett. {\bf{90}}, 052905 {(2007)}.

\bibitem{Shimotani2007}
H. Shimotani, H. Asanuma, A. Tsukazaki, A. Ohtomo, M. Kawasaki, and Y. Iwasa,
Appl. Phys. Lett. {\bf{91}}, 082106 {(2007)}.

\bibitem{Efetov2010}
D. K. Efetov, and P. Kim,
Phys. Rev. Lett. {\bf{105}}, 256805 {(2010)}.

\bibitem{Gonnelli2015}
R. S. Gonnelli, F. Paolucci, E. Piatti, K. Sharda, A. Sola, M. Tortello, J. R. Nair, C. Gerbaldi, M. Bruna, and S. Borini,
Sci. Rep. {\bf{5}}, 09554 {(2015)}.

\bibitem{Dhoot2010}
A. S. Dhoot, S. C. Wimbush, T. Benseman, J. L. MacManus-Driscoll, J. R. Cooper, and R. H. Friend,
Adv. Mater. {\bf{22}}, 2529 {(2010)}.

\bibitem{Ye2009}
J. T. Ye, S. Inoue, K. Kobayashi, Y. Kasahara, H. T. Yuan, H. Shimotani, and Y. Iwasa,
Nat. Mater. {\bf{9}}, 125 {(2009)}.

\bibitem{Sajadi2018}
E. Sajadi, T. Palomaki, Z. Fei, W. Zhao, P. Bement, C. Olsen, S. Luescher, X. Xu, J. A. Folk, and D. H. Cobden,
Science {\bf{362}}, 922 {(2018)}.

\bibitem{Fatemi2018}
V. Fatemi, S. Wu, Y. Cao, L. Brethean, Q. D. Gibson, K. Watanabe, T. Taniguchi, R. J. Cava, and P. Jarrillo-Herrero,
Science {\bf{362}}, 926 {(2018)}.

\bibitem{Ueno2008}
K. Ueno, S. Nakamura, H. Shimotani, A. Ohtomo, N. Kimura, T. Nojima, H. Aoki, Y. Iwasa, and M. Kawasaki,
Nat. Mater. {\bf{7}}, 855 {(2008)}.

\bibitem{Daghero2012}
D. Daghero, F. Paolucci,  A. Sola, M. Tortello, G.~A. Ummarino, M. Agosto, R.~S. Gonnelli, J.~R. Nair, and C. Gerbaldi,
Phys. Rev. Lett. {\bf{108}}, 066807 {(2012)}.

\bibitem{Tortello2013}
M. Tortello, A. Sola, K. Sharda, F. Paolucci, J.~R. Nair, C. Gerbaldi, D. Daghero, R.~S. Gonnelli,
Appl. Surf. Sci. {\bf{269}}, 17 {(2013)}.

\bibitem{Choi2014}
D. Choi, R. Pradheesh, H. Kim, H. Im, Y. Chong, and D.-H. Chae,
Appl. Phys. Lett. {\bf{105}}, 012601 {(2014)}.

\bibitem{Piatti2017}
E. Piatti, D. Daghero, G.~A. Ummarino, F. Laviano, J.~R. Nair, R. Cristiano, A. Casaburi, C. Portesi, A. Sola and R.~S. Gonnelli,
Phys. Rev. B {\bf{95}}, 140501 {(2017)}.

\bibitem{Fiory1990}
A. T. Fiory, A. F. Hebard, R. H. Eick, P. M. Mankiewich, R. E. Howard, and M. L. O'Malle, 
Phys. Rev. Lett. {\bf{65}}, 3441 {(1990)}.

\bibitem{Okamoto1992}
M. Okamoto,  
IEEE Trans. Electron Devices {\bf{39}}, 1661 {(1992)}.

\bibitem{Mannhart1993}
J. Mannhart, J. G. Bednorz, K. A. M\"'uller, D. G. Schlom and J. Str\"obel,
Appl. Phys. Lett. {\bf{62}}, 630 {(1993)}.

\bibitem{Mannhart1993b}
J. Mannhart, J. Str\"obel, J. G. Bednorz, and Ch. Gerber,  
J. Alloys Compd. {\bf{195}}, 519 {(1993)}.

\bibitem{Nishino1989}
T. Nishino, M. Hatano, H. Hasegawa, F. Murai, T. Kure, A. Hiraiwa, K. Yagi, and U. Kawabe,
IEEE Electron Device Lett. {\bf{10}}, 61 {(1989)}.

\bibitem{Bardeen1957}
J. Bardeen, L. N. Cooper, and J. R. Schrieffer,
Phys. Rev. {\bf{106}}, 162 {(1957)}.


\bibitem{DeSimoni2018}
G. De Simoni, F. Paolucci, P. Solinas, E. Strambini, and F. Giazotto,
Nat. Nanotech. {\bf{13}}, 802 {(2018)}.

\bibitem{Varnava2018}
C. Varnava,
Nat. Electron. {\bf{1}}, 374 {(2018)}.

\bibitem{Paolucci2018}
F. Paolucci, G. De Simoni, P. Solinas, E. Strambini, and F. Giazotto,
Nano Lett. {\bf{18}}, 4195 {(2018)}.

\bibitem{Paolucci2019}
F. Paolucci, G. De Simoni, P. Solinas, E. Strambini, N. Ligato, P. Virtanen, A. Braggio, and F. Giazotto,
Phys. Rev. Appl. {\bf{11}}, 024061 {(2019)}.

\bibitem{DeSimoni2019}
G. De Simoni, F. Paolucci, C. Puglia, and F. Giazotto,
ACS Nano {\bf{13}}, 7871 {(2019)}.

\bibitem{Paolucci2019_2}
F. Paolucci, F. Vischi, G. De Simoni, C. Guarcello, P. Solinas, and F. Giazotto,
Nano Lett. {\bf{19}}, 6263 {(2019)}.

\bibitem{Ibach}
H. Ibach, and H. L\"uth,
\textit{Solid-State Physics}, 
(Springer, Berlin, 1995)

\bibitem{Courtois2008}
H. Courtois, M. Meschke, J. T. Peltonen, and J. P. Pekola, 
Phys. Rev. Lett. {\bf{101}}, 067002 {(2008)}.

\bibitem{Ginzburg1950}
V. L. Ginzburg, and L. D. Landau,
Zh. Eksp. Teor. Fiz. {\bf{20}}, 35 {(1950)}.

\bibitem{deGennes}
P.~G. De Gennes,
\textit{Superconductivity Of Metals And Alloys}, 
(Westview Press, Boulder, 1999)

\bibitem{tinkham}
M. Tinkham,
\textit{Introduction to superconductivity}, 
(Courier Dover Publications, New York, 2012)

\bibitem{Morpurgo1998}
A.~F. Morpurgo, T.~M. Klopwijk, and B.~J. van Wees,
Appl. Phys. Lett. {\bf{72}}, 966 {(1998)}.

\bibitem{Holm1932}
R. Holm, and W. Meissner,
Z. Phys. {\bf{74}}, 715 {(1932)}.

\bibitem{Carbotte1990}
J. P. Carbotte,
Rev. Mod. Phys. {\bf{62}}, 1027 {(1990)}.

\bibitem{Fornieri2017}
A. Fornieri, G. Timossi, P. Virtanen, P. Solinas, and F. Giazotto,
Nat. Nanotech. {\bf{12}}, 425 {(2017)}.

\bibitem{Takayanagi1985}
H. Takayanagi, and T. Kawakami,
Phys. Rev. Lett. {\bf{54}}, 2449 {(1985)}.

\bibitem{Akazaki1996}
T. Akazaki, H. Takayanagi, and J. Nitta,
Appl. Phys. Lett. {\bf{68}}, 418 {(1996)}.

\bibitem{Doh2005}
Y-J. Doh, J. A. van Dam, A. L. Roest, E. P. A. M. Bakkers, L. P. Kouwenhoven, and S. De Franceschi, 
Science {\bf{309}}, 272 {(2005)}.

\bibitem{Xiang2006}
J. Xiang, A. Vidan, M. Tinkham, R. M. Westervelt, and C. M. Lieber,  
Nat. Nanotechnol. {\bf{1}}, 208 {(2006)}.

\bibitem{Paajaste2015}
J. Paajaste, M. Amado, S. Roddaro, F. S. Bergeret, D. Ercolani, L. Sorba, and F. Giazotto, 
Nano Lett. {\bf{15}}, 1803 {(2015)}.

\bibitem{Mydosh1965}
J.~A. Mydosh, and H. Meissner, 
Phys. Rev. {\bf{140}}, A1568 {(1965)}.

\bibitem{Harwin2017}
R.~C. Harwin, D.~J. Goldie, and S. Withington,
Supercond. Sci. Technol. {\bf{30}}, 084001 {(2017)}.

\bibitem{Golubov2004}
A.~A. Golubov, M.~Y. Kupriyanov, and E. Il'ichev,
Rev. Mod. Phys. {\bf{76}}, 412 {(2004)}.

\bibitem{Kulik1977}
I.~O. Kulik, and A.~N. Omelyanchuk,
Sov. J. Low Temp. Phys. {\bf{3}}, 945 {(1977)}.

\bibitem{Ambegaokar1963}
V. Ambegaokar, and A. Baratoff, 
Phys. Rev. Lett. {\bf{10}}, 486 {(1963)}.

\bibitem{Larsen2015}
T.~W. Larsen, K.~D. Petersson, F. Kuemmeth, T.~S. Jespersen, P. Krogstrup, J. Nyg\aa rd, and C.~M. Marcus,  
Phys. Rev. Lett. {\bf{115}}, 127001 {(2015)}.

\bibitem{deLange2015}
G. de Lange, B. van Heck, A. Bruno, D.~J. van Woerkom, A. Geresdi, S.~R. Plissard, E.~P.~A.~M Bakkers, A.~R. Akhmerov, and L. DiCarlo, 
Phys. Rev. Lett. {\bf{115}}, 127002 {(2015)}.

\bibitem{Matthias1963}
B. T. Matthias, T. H. Geballe, and V. B. Compton,
Rev. Mod. Phys. {\bf{35}}, 1 {(1963)}.

\bibitem{Giazotto2006}
F. Giazotto, T. T. Heikkil\"a, A. Luukanen, A. M. Savin, and J. P. Pekola, 
Rev. Mod. Phys. {\bf{78}}, 217 {(2006)}.

\bibitem{Savin2004}
A. M. Savin, J. P. Pekola, J. T. Flyktman, A. Anthore, and F. Giazotto,  
Appl. Phys. Lett. {\bf{84}}, 4179 {(2004)}.

\bibitem{Roddaro2011}
S. Roddaro, A. Pescaglini, D. Ercolani, L. Sorba, F. Giazotto, and F. Beltram,
Nano Res. {\bf{4}}, 259 {(2011)}.

\bibitem{Fulton1972}
T. A. Fulton, L. N. Dunkleberger, and R. C. Dynes,
Phys. Rev. B {\bf{6}}, 855 {(1972)}.

\bibitem{Tsang1975}
W.-T. Tsang, and T. Van Duzer.
J. Appl. Phys {\bf{46}}, 4573 {(1975)}.

\bibitem{Annunziata2010}
A. J. Annunziata, D. F. Santavicca, L. Frunzio, G. Catelani, and M. J. Rooks, 
Nanotechnology {\bf{21}}, 445202 {(2010)}.

\bibitem{Barone1982}
A. Barone, and G. Patern\'o,
\textit{Physics and Applications of the Josephson Effect}, 
(Wiley, New York, 1982)

\bibitem{Clarke2004}
J. Clarke, and A.~I. Bragisnki,
\textit{The SQUID Handbook}, 
{(Wiley, Weinheim, 2004)}.

\bibitem{Mooij1999}
J. E. Mooij, T. P. Orlando, L. Levitov, L. Tian, C. H. van der Wal, and S. Lloyd,
Science {\bf{285}}, 1036 {(1999)}.

\bibitem{Yan2016}
F. Yan, S. Gustavsson, A. Kamal, J. Birenbaum, A. P. Sears, D. Hover, T. J. Gudmundsen, D. Rosenberg, G. Samach, S. Weber, J. L. Yoder, T. P. Orlando, J. Clarke, A. J. Kerman, and W. D. Oliver, 
Nat. Commun. {\bf{7}}, 12964 {(2016)}.

\bibitem{Martinis2002}
J. M. Martinis, S. Nam, J. Aumentado, and C. Urbina, 
Phys. Rev. Lett. {\bf{89}}, 117901 {(2002)}.

\bibitem{Koch2007}
J. Koch, T. M. Yu, J. Gambetta, A. A. Houck, D. I. Schuster, J. Majer, A. Blais, M. H. Devoret, S. M. Girvin, and R. J. Schoelkopf,
Phys. Rev. A {\bf{76}}, 042319 {(2007)}.

\bibitem{Schreier2008}
J. A. Schreier, A. A. Houck, J. Koch, D. I. Schuster, B. R. Johnson, J. M. Chow, J. Gambetta, J. Majer, L. Frunzio, M. H. Devoret, S. M. Girvin, and R. J. Schoelkopf,
Phys. Rev. B {\bf{77}}, 180502R {(2008)}.

\bibitem{Barends2013}
R. Barends, J. Kelly, A. Megrant, D. Sank, E. Jeffrey, Y. Chen, Y. Yin, B. Chiaro, J. Mutus, C. Neill, P. OMalley, P. Roushan, J. Wenner, T. C. White, A. N. Cleland, and J. M. Martinis,
Phys. Rev. Lett. {\bf{111}}, 080502 {(2013)}.

\bibitem{Terzioglu1998}
E. Terzioglu, and M. R. Beasley,
IEEE Trans. Appl. Supercond. {\bf{8}}, 48 {(1998)}.

\bibitem{Likharev1995}
K. K. Likharev,
IEEE Trans. Appl. Supercond. {\bf{1}}, 3 {(1995)}.

\bibitem{Worsham1998}
A. H. Worsham, J. X. Przybysz, J. Kang, and D. L. Miller, 
IEEE Trans. Appl. Supercond. {\bf{5}}, 2996 {(1998)}.

\bibitem{Giazotto2010}
F. Giazotto,  J.~T. Peltonen, M. Meschke, and J.~P. Pekola,
Nat. Phys. {\bf{6}}, 254-259 (2010).

\bibitem{Strambini2016}
E. Strambini, S. D'Ambrosio, F. Vischi, F.~S. Bergeret, Yu.~V. Nazarov, and F. Giazotto,
Nat. Nanotech. {\bf{11}}, 1055-1059 (2016).

\bibitem{Giazotto2012}
F. Giazotto, and M.~J. Martinez-P\'erez,
Nature {\bf{492}}, 401-405 (2012).

\bibitem{Mart2014}
M.~J. Martinez-P\'erez, P. Solinas, and F.  Giazotto,
Nature Commun. {\bf{5}}, 3579 (2014).

\bibitem{Perez2014}
M.~J. Martinez-P\'erez, P. Solinas, and F. Giazotto,
J. Low Temp. Phys. {\bf{175}}, 813 {(2014)}.

\bibitem{Fornieri2015}
A. Fornieri, C. Blanc, R. Bosisio, S. D'Ambrosio, and F. Giazotto,
Nat. Nanotech. {\bf{11}}, 258-263 (2015).

\bibitem{Fornieri2017}
A. Fornieri, and F. Giazotto,
Nat. Nanotech. {\bf{12}}, 944 {(2017)}.

\bibitem{Gol2001}
G.~N. Gol'tsman, O. Okunev, G. Chulkova, A. Semenov, K. Smirnov, B. Voronov, A. Dzardanov, and A.Lipatov,
Appl. Phys. Lett. {\bf{79}}, 705 {(2001)}.

\bibitem{Walker1963}
P. A. Walker,
J. I. Electron. Rad. Eng. {\bf{25}}, 387 {(1963)}.

\bibitem{McCaughan2014}
A.~N. McCaughan, and K.~K. Berggren,
Nano Lett. {\bf{14}}, 5784 {(2014)}.

\bibitem{Zhao2017}
Q-Y. Zhao, A.~N. McCaughan, A.~E. Dane, K.~K. Berggren, and T. Ortlepp,
Supercond. Sci. Technol. {\bf{30}}, 044002 {(2017)}.

\bibitem{Wendin2017}
G. Wendin,
Rep. Prog. Phys {\bf{80}}, 106001 {(2017)}.

\bibitem{Likharev2012}
K.~K. Likharev,
Phys. C {\bf{482}}, 6 {(2012)}.

\bibitem{Clark1980}
T.~D. Clark, R.~J. Prance, and A.~D.~C. Grassie,
J. Appl. Phys. {\bf{51}}, 2739 {(1980)}.

\bibitem{Mercaldo2019}
M. T. Mercaldo, P. Solinas, F. Giazotto, and M. Cuoco,
arXiv:1907.09227  {(2019)}.



\end{thebibliography}
\end{document}